\documentclass[runningheads]{llncs}
\usepackage{graphicx}
\usepackage{adjustbox}

\usepackage{hyperref}

\hypersetup{hidelinks,
backref=true,
pagebackref=true,
hyperindex=true,
breaklinks=true,
colorlinks=true,
urlcolor=blue,
bookmarks=true,
bookmarksopen=false,
pdftitle={Title},
pdfauthor={Author}}

\newcommand{\beginsupplement}{%
        \setcounter{section}{0}
        \renewcommand{\thesection}{S\arabic{section}}%
        \setcounter{table}{0}
        \renewcommand{\thetable}{S\arabic{table}}%
        \setcounter{figure}{0}
        \renewcommand{\thefigure}{S\arabic{figure}}%
     }

\usepackage[figuresright]{rotating}
\usepackage[misc,geometry]{ifsym}
\usepackage{ amssymb }
\usepackage[flushleft]{threeparttable}
\usepackage{booktabs}

\usepackage{pdflscape}

\graphicspath{{./figures/}}

\begin{document}
\title{MRI-based classification of IDH mutation and 1p/19q codeletion status of gliomas using a 2.5D hybrid multi-task convolutional neural network}
%
\titlerunning{2.5D hybrid multi-task CNN for molecular subtyping}

\author{
Satrajit Chakrabarty\inst{1}\Letter \and
Pamela LaMontagne\inst{2} \and
Joshua Shimony\inst{2} \and
Daniel S. Marcus\inst{2} \and
Aristeidis Sotiras\inst{2,5}\Letter
}

\authorrunning{S. Chakrabarty et al.}
%
\institute{Washington University in St. Louis, St. Louis, MO 63130, USA \and
Mallinckrodt Institute of Radiology, Washington University School of Medicine, St. Louis, MO 63110, USA \and
Institute for Informatics, Washington University School of Medicine, St. Louis, MO 63110, USA\\
\Letter~Correspondence: \email{\{satrajit.chakrabarty,aristeidis.sotiras\}@wustl.edu}
}

%
\maketitle              
\begin{abstract}
Isocitrate dehydrogenase (IDH) mutation and 1p/19q codeletion status are important prognostic markers for glioma. Currently, they are determined using invasive procedures. Our goal was to develop artificial intelligence-based methods to non-invasively determine these molecular alterations from MRI. For this purpose, pre-operative MRI scans of 2648 patients with gliomas (grade II-IV) were collected from Washington University School of Medicine (WUSM; n = 835) and publicly available datasets viz. Brain Tumor Segmentation (BraTS; n = 378), LGG 1p/19q (n = 159), Ivy Glioblastoma Atlas Project (Ivy GAP; n = 41), The Cancer Genome Atlas (TCGA; n = 461), and the Erasmus Glioma Database (EGD; n = 774). A 2.5D hybrid convolutional neural network was proposed to simultaneously localize the tumor and classify its molecular status by leveraging imaging features from MR scans and prior knowledge features from clinical records and tumor location. The models were tested on one internal (TCGA) and two external (WUSM and EGD) test sets. For IDH, the best-performing model achieved areas under the receiver operating characteristic (AUROC) of 0.925, 0.874, 0.933 and areas under the precision-recall curves (AUPRC) of 0.899, 0.702, 0.853 on the internal, WUSM, and EGD test sets, respectively. For 1p/19q, the best model achieved AUROCs of 0.782, 0.754, 0.842, and AUPRCs of 0.588, 0.713, 0.782, on those three data-splits, respectively. The high accuracy of the model on unseen data showcases its generalization capabilities and suggests its potential to perform a ‘virtual biopsy’ for tailoring treatment planning and overall clinical management of gliomas.

\keywords{deep learning \and glioma \and isocitrate dehydrogenase \and 1p/19q codeletion \and overall survival \and multi-task \and molecular subtyping.}
\end{abstract}

\section{Introduction}
Gliomas are characterized by distinct imaging characteristics, response to therapy, prognoses, and varying survival rate. As per the 2016 World Health Organization (WHO) guidelines~\cite{louis20162016}, the definition of these tumors requires integrating histological information with molecular parameters. Two of the most important molecular markers are the mutation status of isocitrate dehydrogenase (IDH) enzyme and the codeletion of chromosome arms 1p and 19q (1p/19q). These markers have unique prognostic significance that can considerably impact treatment planning. Therefore, an accurate determination of these molecular parameters can significantly improve patient outcome. 

The current clinical gold-standard for identifying IDH and 1p/19q status involves invasive brain biopsy procedures that can have associated risk~\cite{jackson2001limitations}, may fail to capture intra-tumoral spatial heterogeneity, can be inaccessible in low-resource settings, or can lack adequate tumor content or optimal quality and quantity of nucleic acid required for correct molecular characterization~\cite{cancer2008comprehensive}. Therefore, non-invasive imaging techniques, like MRI, have been investigated as complementary ‘virtual biopsy’ procedures.

Artificial intelligence-based approaches~\cite{korfiatis2019deep} have attempted to perform molecular assessment by leveraging the variation in tumor phenotypical characteristics manifested in MRI scans due to changes in molecular alterations~\cite{chang2018deep,rathore2020multi,batchala2019neuroimaging,shboul2020prediction,van2019predicting}. Several studies~\cite{van2021accuracy,bhandari2021noninvasive} have investigated machine learning (ML) approaches in conjunction with radiomic features for this purpose. However, these methods are limited by their reliance on manually engineered features, lack of standardized methods for feature selection, and reproducibility issues associated with radiomic features~\cite{traverso2018repeatability}. On the other hand, deep learning (DL) approaches~\cite{korfiatis2019deep} overcome these limitations by automatically learning hierarchical imaging features. Nevertheless, several challenges still limit their adoption in routine clinical practice. First, most methods require a manually drawn~\cite{matsui2020prediction,chang2018residual,pasquini2021deep}, or automatically generated~\cite{chang2018deep,li2017deep,decuyper2021automated,choi2021fully} segmentation mask of the tumor. Manual delineation of tumor masks is tedious and prone to human error and observer bias, whereas automatically generated masks require an additional task-specific model. Such task-specific models not only increase computational burden, but also fail to leverage the context between different related tasks. To address this, multi-task DL models have been proposed~\cite{decuyper2021automated,van2022combined}. However, these focus solely on imaging information and fail to incorporate prior clinical knowledge.  Second, most studies have assessed their methods either on one type of molecular status (only IDH mutation~\cite{chang2018residual,park2021mri} or 1p/19q codeletion~\cite{van2019predicting}) or specific grades of glioma (e.g., only low grade~\cite{matsui2020prediction,li2017deep}). This failed to provide a comprehensive classification system that aligns with WHO 2016 classification and recognizes the importance of combined IDH and 1p/19q status prediction. Third, previous studies often used small samples and lacked rigorous external validation~\cite{matsui2020prediction,pasquini2021deep}, which is necessary for accurately assessing model generalizability. Fourth, existing studies have used varying datasets and performance metrics that make objective comparisons between various methods challenging. Without head-to-head comparisons and data-driven conclusions, it is difficult to gauge the advancements in the field and identify the best-performing methods. 

To address these limitations, we propose a 2.5D multi-task hybrid convolutional neural network (CNN) approach for classifying both IDH mutation and 1p/19q codeletion status of high- and low-grade gliomas (grade II-IV) from routine MR sequences (i.e., pre-operative post-contrast T1-weighted (T1c), T2-weighted (T2), and T2-weighted Fluid-attenuated inversion recovery (FLAIR)). Our model jointly detects and segments the glioma before classifying its molecular status, thus obviating any additional tumor segmentation step. Additionally, it can integrate prior knowledge through a feature-fusion mechanism. We train the model on three orthogonal planes viz. axial, coronal, and sagittal, thus providing the model with volumetric context information without incurring the computational burden of a 3D model. We assembled the largest sample till date for a study of this kind, consisting of 2648 patients from 14 institutions. The model has been extensively validated on three independent hold-out sets comprising 968 patient cases from 11 different institutions, to demonstrate its generalizability.

\section{Materials and methods}
Retrospective deidentified data was obtained from WUSM, with a waiver of consent in accordance with the Health Insurance Portability and Accountability Act, as approved by the Institutional Review Board (IRB ID \# 202004209). Additional data was obtained from public datasets after completion of necessary data usage agreements.

\begin{figure}
\centering
\includegraphics[width=\textwidth]{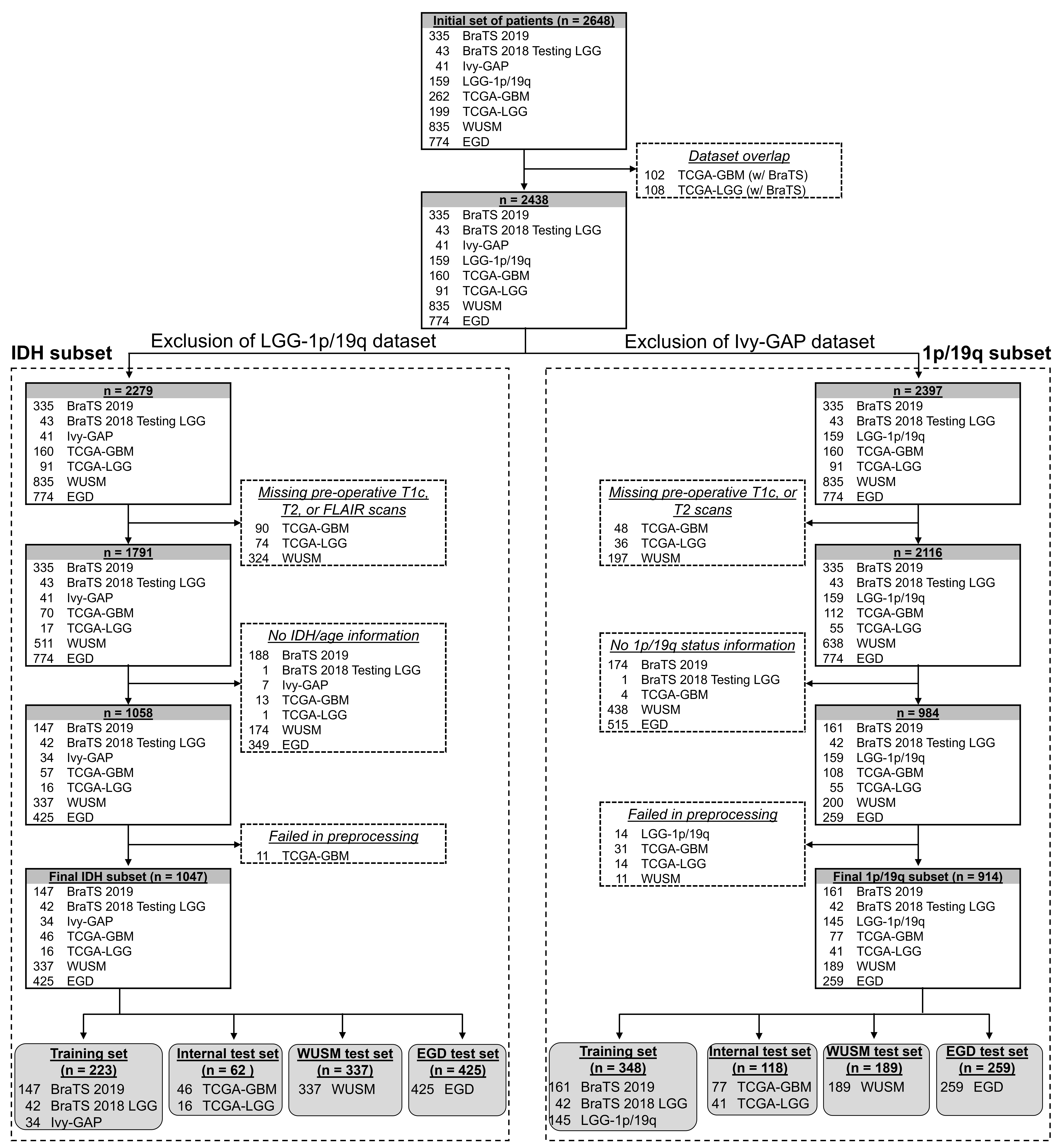}
\caption{Inclusion flow-chart of data for IDH mutation and 1p/19q codeletion experiments. The underlined and italicized text within each box with dashed lines details the reason for exclusions. BraTS: Brain Tumor Segmentation dataset, Ivy-GAP: Ivy Glioblastoma Atlas Project, TCGA-GBM: The Cancer Genome Atlas Glioblastoma Multiforme, TCGA-LGG: The Cancer Genome Atlas Low Grade Glioma, WUSM: Washington University School of Medicine, EGD: Erasmus Glioma Database, T1c: post-contrast T1-weighted scan, T2: T2-weighted scan, FLAIR: Fluid Attenutated Inversion Recovery scan.} 
\label{fig1}
\end{figure}

\subsection{Datasets}
Retrospective pre-operative MRI scans from 2648 patients with histopathologically confirmed gliomas (grade II-IV) were considered for inclusion in the study (Figure~\ref*{fig1}). Data was acquired from seven publicly available datasets across 13 different institutions: Brain Tumor Segmentation (BraTS)~\cite{menze2014multimodal,bakas2017advancing,bakas2018identifying} 2019 (n = 335) and 2018 (n = 43), LGG 1p/19q~\cite{erickson2017data} (n = 159), Ivy Glioblastoma Atlas Project (Ivy GAP)~\cite{puchalski2018anatomic} (n = 41), The Cancer Genome Atlas Glioblastoma Multiforme (TCGA-GBM)~\cite{bakas2017segmentationgbm} (n = 262), TCGA-Low Grade Glioma (TCGA-LGG)~\cite{bakas2017segmentationlgg} (n = 199), and the Erasmus Glioma Database (EGD)~\cite{van2021erasmus} (n = 774). Additional patient cases were acquired from retrospective health records of Washington University School of Medicine (WUSM; n = 835). No patient cases were excluded based on image acquisition parameters or image quality to mirror the inherent heterogeneity present in clinical data. Additionally, to ensure a wide applicability of the method, only routine MRI sequences were used, and no exclusions were made based on glioma grade. The BraTS 2019, 2018, LGG-1p/19q, Ivy GAP, and EGD datasets included expert-annotated multi-class tumor segmentation masks. These comprise edema, non-enhancing/necrotic tumor core and enhancing tumor, which were combined to define the whole tumor class. The WUSM dataset included overall survival (OS) information of patients. No OS information was available for the EGD dataset and hence survival analysis could not be performed for this data. 

Based on study requirements, two different but overlapping subsets of data were considered for IDH and 1p/19q status classification. For both classification tasks, the training sets included only cases with available expert tumor segmentations, required to train the model. An internal test set, and two additional external sets were included for each task to accurately estimate model generalizability. 

For compilation of the IDH database, the inclusion criteria were: (i) pathologically confirmed glioma (grade II-IV), (ii) known IDH status (see Supplementary data~\ref*{supp_methods_gt}) and patient age at diagnosis, and (iii) presence of T1c, T2, and FLAIR scans. Based on these criteria, 1047 patient cases were selected. These were subsequently split into four sets: cross-validation (n = 223 from BraTS 2019, 2018, and Ivy GAP), internal testing (n = 62 from TCGA-GBM and TCGA-LGG), and two external test sets viz. WUSM (WUSM test set; n = 337) and EGD (EGD test set; n = 425).

For compilation of the 1p/19q database, the following inclusion criteria were considered: (i) pathologically confirmed glioma (grade II-IV), (ii) known 1p/19q codeletion status (see Supplementary data~\ref*{supp_methods_gt}), and (iii) presence of T1c and T2 scans. We did not require the presence of FLAIR because the LGG 1p/19q dataset comprised only T1c and T2 scans. Based on these criteria, 914 patient cases were selected. These were subsequently split into four sets: cross-validation (n = 348 from BraTS 2019, 2018, and LGG 1p/19q), internal testing (n = 118 from TCGA-GBM and TCGA-LGG), and two external test sets viz. WUSM (WUSM test set; n = 189) and EGD (EGD test set; n = 259).

\subsection{Image acquisition, pre-processing, and feature extraction}
Due to being acquired from eight different sources across 14 different institutions, the data was extremely heterogeneous exhibiting high variability in acquisition protocol parameters (see Supplementary data~\ref*{supp_methods_dataacq}, Supplementary Figures~\ref*{supp_fig1}-\ref*{supp_fig5}). All data were pre-processed in a standardized way using the Multimodal Glioma Analysis pipeline~\cite{chakrabarty2020preprocessing} (registration to a common anatomical atlas~\cite{rohlfing2010sri24}, skull-stripping and intensity normalization; see Supplementary data~\ref*{supp_methods_preproc}).

Besides imaging data, our network can integrate prior clinical knowledge by incorporating additional features in a late-fusion step. To this end, two prior knowledge features viz. patient age at diagnosis (hereon referred to as ‘age’) and anatomical location of tumor (hereon referred to as ‘loc’) were included in the network (see Supplementary data~\ref*{supp_methods_preproc}).

\begin{figure}
\centering
\includegraphics[width=\textwidth]{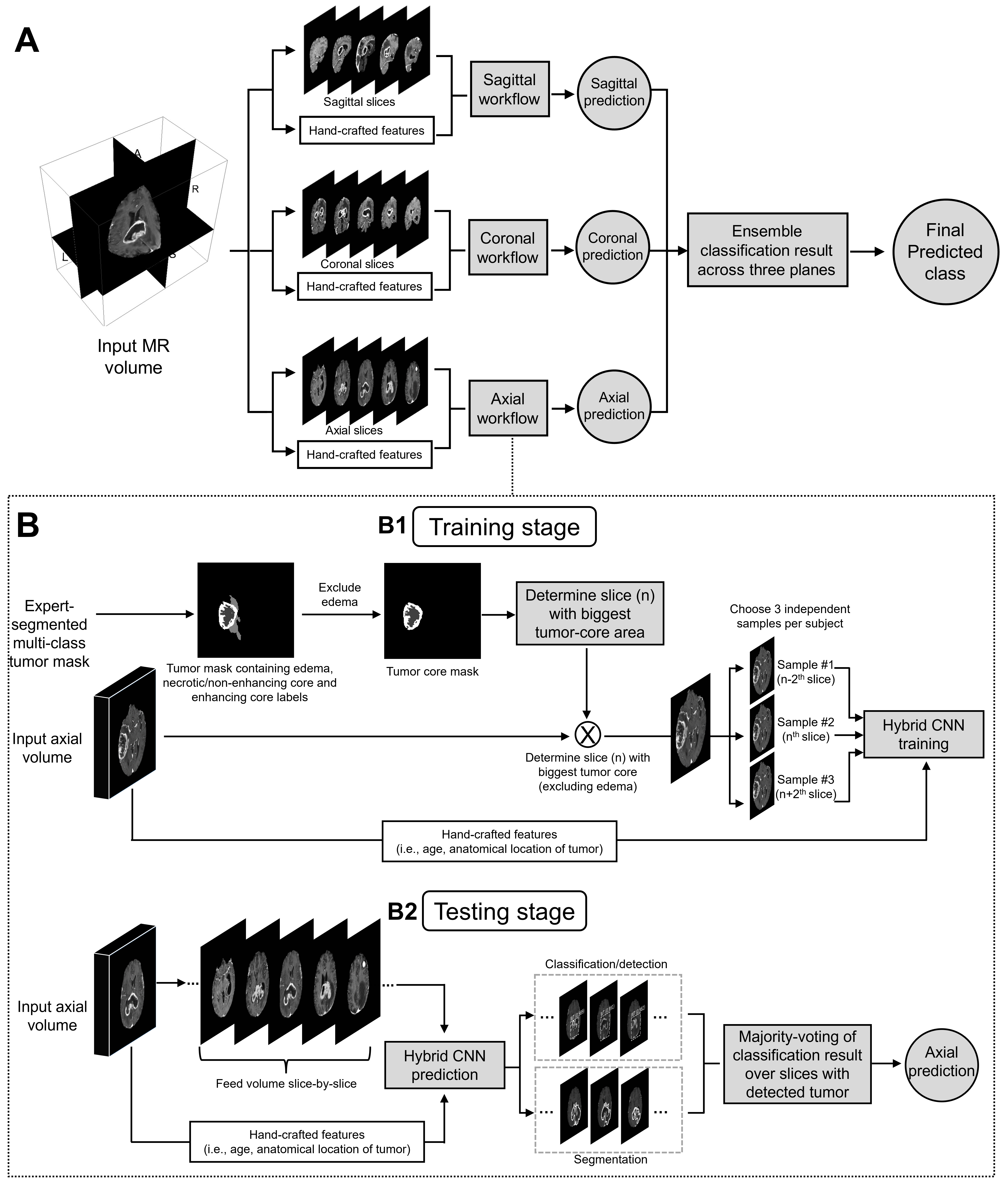}
\caption{Detailed schematic of the proposed deep learning system for predicting molecular status. (A) Multi-view aggregation architecture that combines predictions from planar models trained independently on 2D slices along three principal axes (axial, coronal, sagittal). The final classification result is obtained by aggregating the classification result from all three networks. (B) Each planar 2D network is based on a hybrid network that simultaneously detects and segments the tumor and classifies the molecular status, based on 2D slices and prior knowledge features. During training (subpanel B1) expert-segmented multi-class tumor masks are used to i) select 2D slices and ii) supervise the tumor detection and segmentation tasks. During testing (subpanel B2), the 2D planar model is applied to all available slices and a prediction is made through consensus of predictions for slices where a tumor was detected. For ease of visualization, we show only the axial workflow for one modality (post-contrast T1-weighted), but models work across three planes for multiple modalities. MR: Magnetic Resonance, CNN: Convolutional Neural Network.} 
\label{fig2}
\end{figure}

\subsection{Hybrid 2.5D multi-task model architecture}
We adopted a 2.5D approach, aiming to capture volumetric spatial information, while minimizing computational requirements. Specifically, we train a separate 2D model for each of the three orthogonal planes (i.e., axial, coronal, and sagittal), whose predictions are combined into the final result through a multi-view aggregation step (Figure~\ref*{fig2}). 

The end goal of each of these 2D models is to classify the molecular status. However, due to the sparse presence of glioma in the MRI image, the classification performance of the network might get affected by non-tumorous image characteristics. To resolve this, the proposed 2D models follow a Mask RCNN architecture~\cite{he2017mask} and tackle two auxiliary tasks of detection and segmentation of the tumor besides the classification task (Supplementary data~\ref*{supp_methods_model}, Supplementary Figure~\ref*{supp_fig6}). Additionally, we augment the 2D models into a hybrid architecture, which integrates imaging features with prior knowledge features. 

CNNs are mostly image-intensity based, and hence are unable to take demographic features (e.g., patient age, etc.) or neuroanatomical features (e.g., tumor location, etc.) into account. This is limiting given ample evidence regarding the importance of patient age in predicting IDH status~\cite{zhang2017multimodal} and the association between tumor location and 1p/19q codeleted status~\cite{chang2018deep,shboul2020prediction,van2019predicting}. To tackle this limitation, we equipped our CNN architecture with the ability to exploit additional features (i.e., ‘age’ and ‘loc’ features), thus combining the strengths of data-driven features with clinical prior knowledge. To incorporate these features into the network, we implemented a late-fusion strategy~\cite{zhang2019late}, where the additional features are normalized and concatenated with the CNN feature output in the classification/detection head (Supplementary Figure~\ref*{supp_fig6}A). Subsequently, this set of hybrid features is passed through a fully connected (FC) layer to the final classification layer of the network. The training and testing processes involving the hybrid features are end-to-end (Figure~\ref*{fig2}, Supplementary data~\ref*{supp_methods_training}).

\subsection{Statistical analysis}
We used Chi-square and Mann-Whitney tests to evaluate differences in patient demographics and clinical characteristics between data splits. We performed ablation studies to determine the importance of prior knowledge features and to investigate the effectiveness of 2.5D compared to 2D models (see Supplementary data~\ref*{supp_methods_ablation}). The performance of the best-performing model was compared to two baseline pre-trained models: (i) a multi-task U-net model by Voort et al.~\cite{van2022combined} (hereon referred to as “Voort-CNN”) for both IDH and 1p/19q prediction tasks, and (ii) a CNN-radiomics hybrid model by Choi et al.~\cite{choi2021fully} (hereon referred to as “Choi-CNN”) for only the IDH prediction task (see Supplementary data~\ref*{supp_methods_baselines}).

The classification performance was quantified using accuracy, precision, recall, F1 score, area under receiver operating characteristics (AUROC), and area under precision-recall curves (AUPRC). For AUROC and AUPRC, 95\% confidence intervals (CI) were calculated using a 1000-sample bootstrapping method (see Supplementary data~\ref*{supp_methods_stat}). Confusion matrices were calculated to show the error distribution across different classes. Statistical comparisons between methods were performed using the McNemar test~\cite{mcnemar1947note} for precision, the generalized score statistic~\cite{leisenring2000comparisons} for recall, and the DeLong test~\cite{delong1988comparing} for AUROCs.

To assess the validity of our model in terms of WHO 2016 glioma subtypes, we compared overall survival (OS) based on ground truth molecular status to OS based on predicted status (see Supplementary data~\ref*{supp_methods_survival}). We hypothesized that the misclassified IDH-wt cases with IDH-mut like phenotype will have better OS. Accordingly, we used Kaplan-Meier survival curves to characterize and compare groups of misclassified cases (i.e., IDH-wt predicted as IDH-mut and vice versa) in WUSM in terms of OS. Differences in the Kaplan-Meier survival curves were analyzed using Cox regression analysis. Additionally, we examined recurring patterns in misclassified cases for both IDH and 1p/19q classification tasks (see Supplementary data~\ref*{supp_methods_survival}).

\begin{landscape}
\begin{table}[]
\scalebox{0.9}{
\begin{threeparttable}
\caption{Patient characteristics stratified by train, test, WUSM, and EGD splits of data.}
\label{table1}
\begin{tabular}{l|ccccccc|ccccccc}
\toprule
              & \multicolumn{7}{c|}{IDH}                                                                                                                                                                                                                                                                                                                                                                                                                                                                                                                             & \multicolumn{7}{c}{1p/19q}                                                                                                                                                                                                                                                                                                                                                                                                                                                                                                                           \\ \hline
              & \multicolumn{1}{c|}{\begin{tabular}[c]{@{}c@{}}train\\ (n = 223)\end{tabular}} & \multicolumn{1}{c|}{\begin{tabular}[c]{@{}c@{}}test\\ (n = 62)\end{tabular}} & \multicolumn{1}{c|}{\begin{tabular}[c]{@{}c@{}}P-value\\ (test)\end{tabular}} & \multicolumn{1}{c|}{\begin{tabular}[c]{@{}c@{}}ext\\ (n = 337)\end{tabular}} & \multicolumn{1}{c|}{\begin{tabular}[c]{@{}c@{}}P-value\\ (ext)\end{tabular}} & \multicolumn{1}{c|}{\begin{tabular}[c]{@{}c@{}}EGD\\ (n = 425)\end{tabular}} & \begin{tabular}[c]{@{}c@{}}P-value\\ (EGD)\end{tabular} & \multicolumn{1}{c|}{\begin{tabular}[c]{@{}c@{}}train\\ (n = 348)\end{tabular}} & \multicolumn{1}{c|}{\begin{tabular}[c]{@{}c@{}}test\\ (n = 117)\end{tabular}} & \multicolumn{1}{c|}{\begin{tabular}[c]{@{}c@{}}P-value\\ (test)\end{tabular}} & \multicolumn{1}{c|}{\begin{tabular}[c]{@{}c@{}}ext\\ (n = 189)\end{tabular}} & \multicolumn{1}{c|}{\begin{tabular}[c]{@{}c@{}}P-value\\ (ext)\end{tabular}} & \multicolumn{1}{c|}{\begin{tabular}[c]{@{}c@{}}EGD\\ (n = 248)\end{tabular}} & \begin{tabular}[c]{@{}c@{}}P-value\\ (EGD)\end{tabular} \\ \midrule
Age           & \multicolumn{1}{c|}{\begin{tabular}[c]{@{}c@{}}54\\ (41 - 62)\end{tabular}}    & \multicolumn{1}{c|}{\begin{tabular}[c]{@{}c@{}}58\\ (49 - 66)\end{tabular}}  & \multicolumn{1}{c|}{0.041}                                                    & \multicolumn{1}{c|}{\begin{tabular}[c]{@{}c@{}}59\\ (46 - 68)\end{tabular}}  & \multicolumn{1}{c|}{\textless{}0.001}                                        & \multicolumn{1}{c|}{\begin{tabular}[c]{@{}c@{}}57\\ (45 - 69)\end{tabular}}  & 0.0017                                                  & \multicolumn{1}{c|}{\begin{tabular}[c]{@{}c@{}}48\\ (35 - 59)\end{tabular}}    & \multicolumn{1}{c|}{\begin{tabular}[c]{@{}c@{}}56\\ (38 - 66)\end{tabular}}   & \multicolumn{1}{c|}{0.001}                                                    & \multicolumn{1}{c|}{\begin{tabular}[c]{@{}c@{}}48\\ (33 - 62)\end{tabular}}  & \multicolumn{1}{c|}{0.8282}                                                  & \multicolumn{1}{c|}{\begin{tabular}[c]{@{}c@{}}48\\ (37 - 58)\end{tabular}}  & 0.8613                                                  \\ \hline
Sex           & \multicolumn{1}{c|}{}                                                          & \multicolumn{1}{c|}{}                                                        & \multicolumn{1}{c|}{0.4838}                                                   & \multicolumn{1}{c|}{}                                                        & \multicolumn{1}{c|}{0.0406}                                                  & \multicolumn{1}{c|}{}                                                        & 0.0265                                                  & \multicolumn{1}{c|}{}                                                          & \multicolumn{1}{c|}{}                                                         & \multicolumn{1}{c|}{0.9323}                                                   & \multicolumn{1}{c|}{}                                                        & \multicolumn{1}{c|}{0.004}                                                   & \multicolumn{1}{c|}{}                                                        & 0.0107                                                  \\ 
~~~~~Female        & \multicolumn{1}{c|}{\begin{tabular}[c]{@{}c@{}}107\\ (47.98\%)\end{tabular}}   & \multicolumn{1}{c|}{\begin{tabular}[c]{@{}c@{}}26\\ (41.94\%)\end{tabular}}  & \multicolumn{1}{c|}{}                                                         & \multicolumn{1}{c|}{\begin{tabular}[c]{@{}c@{}}131\\ (38.87\%)\end{tabular}} & \multicolumn{1}{c|}{}                                                        & \multicolumn{1}{c|}{\begin{tabular}[c]{@{}c@{}}164\\ (38.59\%)\end{tabular}} &                                                         & \multicolumn{1}{c|}{\begin{tabular}[c]{@{}c@{}}164\\ (47.13\%)\end{tabular}}   & \multicolumn{1}{c|}{\begin{tabular}[c]{@{}c@{}}55\\ (47.01\%)\end{tabular}}   & \multicolumn{1}{c|}{}                                                         & \multicolumn{1}{c|}{\begin{tabular}[c]{@{}c@{}}64\\ (33.86\%)\end{tabular}}  & \multicolumn{1}{c|}{}                                                        & \multicolumn{1}{c|}{\begin{tabular}[c]{@{}c@{}}90\\ (36.29\%)\end{tabular}}  &                                                         \\ 
~~~~~Male          & \multicolumn{1}{c|}{\begin{tabular}[c]{@{}c@{}}116\\ (52.02\%)\end{tabular}}   & \multicolumn{1}{c|}{\begin{tabular}[c]{@{}c@{}}36\\ (58.06\%)\end{tabular}}  & \multicolumn{1}{c|}{}                                                         & \multicolumn{1}{c|}{\begin{tabular}[c]{@{}c@{}}206\\ (61.13\%)\end{tabular}} & \multicolumn{1}{c|}{}                                                        & \multicolumn{1}{c|}{\begin{tabular}[c]{@{}c@{}}261\\ (61.41\%)\end{tabular}} &                                                         & \multicolumn{1}{c|}{\begin{tabular}[c]{@{}c@{}}184\\ (52.87\%)\end{tabular}}   & \multicolumn{1}{c|}{\begin{tabular}[c]{@{}c@{}}62\\ (52.99\%)\end{tabular}}   & \multicolumn{1}{c|}{}                                                         & \multicolumn{1}{c|}{\begin{tabular}[c]{@{}c@{}}125\\ (66.14\%)\end{tabular}} & \multicolumn{1}{c|}{}                                                        & \multicolumn{1}{c|}{\begin{tabular}[c]{@{}c@{}}158\\ (63.71\%)\end{tabular}} &                                                         \\ \hline
IDH status    & \multicolumn{1}{c|}{}                                                          & \multicolumn{1}{c|}{}                                                        & \multicolumn{1}{c|}{0.2829}                                                   & \multicolumn{1}{c|}{}                                                        & \multicolumn{1}{c|}{\textless{}0.001}                                        & \multicolumn{1}{c|}{}                                                        & 0.1757                                                  & \multicolumn{1}{c|}{}                                                          & \multicolumn{1}{c|}{}                                                         & \multicolumn{1}{c|}{}                                                         & \multicolumn{1}{c|}{}                                                        & \multicolumn{1}{c|}{}                                                        & \multicolumn{1}{c|}{}                                                        &                                                         \\ 
~~~~~Mutant        & \multicolumn{1}{c|}{\begin{tabular}[c]{@{}c@{}}91\\ (40.81\%)\end{tabular}}    & \multicolumn{1}{c|}{\begin{tabular}[c]{@{}c@{}}20\\ (32.26\%)\end{tabular}}  & \multicolumn{1}{c|}{}                                                         & \multicolumn{1}{c|}{\begin{tabular}[c]{@{}c@{}}88\\ (26.11\%)\end{tabular}}  & \multicolumn{1}{c|}{}                                                        & \multicolumn{1}{c|}{\begin{tabular}[c]{@{}c@{}}149\\ (35.06\%)\end{tabular}} &                                                         & \multicolumn{1}{c|}{-}                                                         & \multicolumn{1}{c|}{-}                                                        & \multicolumn{1}{c|}{}                                                         & \multicolumn{1}{c|}{-}                                                       & \multicolumn{1}{c|}{}                                                        & \multicolumn{1}{c|}{-}                                                       &                                                         \\ 
~~~~~Wildtype      & \multicolumn{1}{c|}{\begin{tabular}[c]{@{}c@{}}132\\ (59.19\%)\end{tabular}}   & \multicolumn{1}{c|}{\begin{tabular}[c]{@{}c@{}}42\\ (67.74\%)\end{tabular}}  & \multicolumn{1}{c|}{}                                                         & \multicolumn{1}{c|}{\begin{tabular}[c]{@{}c@{}}249\\ (73.89\%)\end{tabular}} & \multicolumn{1}{c|}{}                                                        & \multicolumn{1}{c|}{\begin{tabular}[c]{@{}c@{}}276\\ (64.94\%)\end{tabular}} &                                                         & \multicolumn{1}{c|}{-}                                                         & \multicolumn{1}{c|}{-}                                                        & \multicolumn{1}{c|}{}                                                         & \multicolumn{1}{c|}{-}                                                       & \multicolumn{1}{c|}{}                                                        & \multicolumn{1}{c|}{-}                                                       &                                                         \\ \hline
1p/19q        & \multicolumn{1}{c|}{}                                                          & \multicolumn{1}{c|}{}                                                        & \multicolumn{1}{c|}{}                                                         & \multicolumn{1}{c|}{}                                                        & \multicolumn{1}{c|}{}                                                        & \multicolumn{1}{c|}{}                                                        &                                                         & \multicolumn{1}{c|}{}                                                          & \multicolumn{1}{c|}{}                                                         & \multicolumn{1}{c|}{\textless{}0.001}                                         & \multicolumn{1}{c|}{}                                                        & \multicolumn{1}{c|}{0.9945}                                                  & \multicolumn{1}{c|}{}                                                        & 0.1356                                                  \\ 
~~~~~Codeleted     & \multicolumn{1}{c|}{-}                                                         & \multicolumn{1}{c|}{-}                                                       & \multicolumn{1}{c|}{}                                                         & \multicolumn{1}{c|}{-}                                                       & \multicolumn{1}{c|}{}                                                        & \multicolumn{1}{c|}{-}                                                       &                                                         & \multicolumn{1}{c|}{\begin{tabular}[c]{@{}c@{}}121\\ (34.77\%)\end{tabular}}   & \multicolumn{1}{c|}{\begin{tabular}[c]{@{}c@{}}16\\ (13.68\%)\end{tabular}}   & \multicolumn{1}{c|}{}                                                         & \multicolumn{1}{c|}{\begin{tabular}[c]{@{}c@{}}65\\ (34.39\%)\end{tabular}}  & \multicolumn{1}{c|}{}                                                        & \multicolumn{1}{c|}{\begin{tabular}[c]{@{}c@{}}71\\ (28.63\%)\end{tabular}}  &                                                         \\ 
~~~~~Non-codeleted & \multicolumn{1}{c|}{-}                                                         & \multicolumn{1}{c|}{-}                                                       & \multicolumn{1}{c|}{}                                                         & \multicolumn{1}{c|}{-}                                                       & \multicolumn{1}{c|}{}                                                        & \multicolumn{1}{c|}{-}                                                       &                                                         & \multicolumn{1}{c|}{\begin{tabular}[c]{@{}c@{}}227\\ (65.23\%)\end{tabular}}   & \multicolumn{1}{c|}{\begin{tabular}[c]{@{}c@{}}101\\ (86.32\%)\end{tabular}}  & \multicolumn{1}{c|}{}                                                         & \multicolumn{1}{c|}{\begin{tabular}[c]{@{}c@{}}124\\ (65.61\%)\end{tabular}} & \multicolumn{1}{c|}{}                                                        & \multicolumn{1}{c|}{\begin{tabular}[c]{@{}c@{}}177\\ (71.37\%)\end{tabular}} &                                                         \\ \hline
WHO Grade     & \multicolumn{1}{c|}{}                                                          & \multicolumn{1}{c|}{}                                                        & \multicolumn{1}{c|}{\textless{}0.001}                                         & \multicolumn{1}{c|}{}                                                        & \multicolumn{1}{c|}{\textless{}0.001}                                        & \multicolumn{1}{c|}{}                                                        & \textless{}0.001                                        & \multicolumn{1}{c|}{}                                                          & \multicolumn{1}{c|}{}                                                         & \multicolumn{1}{c|}{\textless{}0.001}                                         & \multicolumn{1}{c|}{}                                                        & \multicolumn{1}{c|}{\textless{}0.001}                                        & \multicolumn{1}{c|}{}                                                        & \textless{}0.001                                        \\ 
~~~~~II            & \multicolumn{1}{c|}{\begin{tabular}[c]{@{}c@{}}47\\ (21.08\%)\end{tabular}}    & \multicolumn{1}{c|}{\begin{tabular}[c]{@{}c@{}}11\\ (17.74\%)\end{tabular}}  & \multicolumn{1}{c|}{}                                                         & \multicolumn{1}{c|}{\begin{tabular}[c]{@{}c@{}}66\\ (19.58\%)\end{tabular}}  & \multicolumn{1}{c|}{}                                                        & \multicolumn{1}{c|}{\begin{tabular}[c]{@{}c@{}}124\\ (29.18\%)\end{tabular}} &                                                         & \multicolumn{1}{c|}{\begin{tabular}[c]{@{}c@{}}143\\ (41.09\%)\end{tabular}}   & \multicolumn{1}{c|}{\begin{tabular}[c]{@{}c@{}}28\\ (23.93\%)\end{tabular}}   & \multicolumn{1}{c|}{}                                                         & \multicolumn{1}{c|}{\begin{tabular}[c]{@{}c@{}}91\\ (48.15\%)\end{tabular}}  & \multicolumn{1}{c|}{}                                                        & \multicolumn{1}{c|}{\begin{tabular}[c]{@{}c@{}}126\\ (50.81\%)\end{tabular}} &                                                         \\ 
~~~~~III           & \multicolumn{1}{c|}{\begin{tabular}[c]{@{}c@{}}59\\ (26.46\%)\end{tabular}}    & \multicolumn{1}{c|}{\begin{tabular}[c]{@{}c@{}}5\\ (8.06\%)\end{tabular}}    & \multicolumn{1}{c|}{}                                                         & \multicolumn{1}{c|}{\begin{tabular}[c]{@{}c@{}}47\\ (13.95\%)\end{tabular}}  & \multicolumn{1}{c|}{}                                                        & \multicolumn{1}{c|}{\begin{tabular}[c]{@{}c@{}}29\\ (6.82\%)\end{tabular}}   &                                                         & \multicolumn{1}{c|}{\begin{tabular}[c]{@{}c@{}}109\\ (31.32\%)\end{tabular}}   & \multicolumn{1}{c|}{\begin{tabular}[c]{@{}c@{}}13\\ (11.11\%)\end{tabular}}   & \multicolumn{1}{c|}{}                                                         & \multicolumn{1}{c|}{\begin{tabular}[c]{@{}c@{}}26\\ (13.76\%)\end{tabular}}  & \multicolumn{1}{c|}{}                                                        & \multicolumn{1}{c|}{\begin{tabular}[c]{@{}c@{}}39\\ (15.73\%)\end{tabular}}  &                                                         \\ 
~~~~~IV            & \multicolumn{1}{c|}{\begin{tabular}[c]{@{}c@{}}104\\ (46.64\%)\end{tabular}}   & \multicolumn{1}{c|}{\begin{tabular}[c]{@{}c@{}}46\\ (74.19\%)\end{tabular}}  & \multicolumn{1}{c|}{}                                                         & \multicolumn{1}{c|}{\begin{tabular}[c]{@{}c@{}}222\\ (65.88\%)\end{tabular}} & \multicolumn{1}{c|}{}                                                        & \multicolumn{1}{c|}{\begin{tabular}[c]{@{}c@{}}256\\ (60.24\%)\end{tabular}} &                                                         & \multicolumn{1}{c|}{\begin{tabular}[c]{@{}c@{}}96\\ (27.59\%)\end{tabular}}    & \multicolumn{1}{c|}{\begin{tabular}[c]{@{}c@{}}76\\ (64.96\%)\end{tabular}}   & \multicolumn{1}{c|}{}                                                         & \multicolumn{1}{c|}{\begin{tabular}[c]{@{}c@{}}72\\ (38.1\%)\end{tabular}}   & \multicolumn{1}{c|}{}                                                        & \multicolumn{1}{c|}{\begin{tabular}[c]{@{}c@{}}65\\ (26.21\%)\end{tabular}}  &                                                         \\ 
~~~~~Unknown       & \multicolumn{1}{c|}{\begin{tabular}[c]{@{}c@{}}13\\ (5.83\%)\end{tabular}}     & \multicolumn{1}{c|}{}                                                        & \multicolumn{1}{c|}{}                                                         & \multicolumn{1}{c|}{\begin{tabular}[c]{@{}c@{}}2\\ (0.59\%)\end{tabular}}    & \multicolumn{1}{c|}{}                                                        & \multicolumn{1}{c|}{\begin{tabular}[c]{@{}c@{}}16\\ (3.76\%)\end{tabular}}   &                                                         & \multicolumn{1}{c|}{}                                                          & \multicolumn{1}{c|}{}                                                         & \multicolumn{1}{c|}{}                                                         & \multicolumn{1}{c|}{}                                                        & \multicolumn{1}{c|}{}                                                        & \multicolumn{1}{c|}{\begin{tabular}[c]{@{}c@{}}18\\ (7.26\%)\end{tabular}}   &                                                         \\ \bottomrule
\end{tabular}
\begin{tablenotes}
      \small
      \item Percentage values in parentheses within the table represent the proportion of each subgroup within a specific dataset (i.e., train, test, WUSM, EGD). 
      \item For ‘Age’ row, interquartile range (IQR) is given in parentheses. 
      \item P-values demonstrate differences between training data vs. the dataset mentioned in parentheses in header row.
      \item Abbreviations: WUSM, Washington University School of Medicine; EGD, Erasmus Glioma Database.
    \end{tablenotes}
  \end{threeparttable}
}
\end{table}
\end{landscape}

\section{Results}
\subsection{Dataset Characteristics}
Patient demographics and clinical characteristics have been calculated for all datasets created for each prediction task (Table~\ref*{table1}). For both IDH and 1p/19q subsets, the internal, WUSM, and EGD sets differed to varying degrees in terms of age, sex, and clinical characteristics (see Supplementary data~\ref*{supp_results_dataset}).

\subsection{IDH mutation status prediction}
\subsubsection{Classification performance}
\begin{figure}[htbp]
\centering
\includegraphics[width=\textwidth]{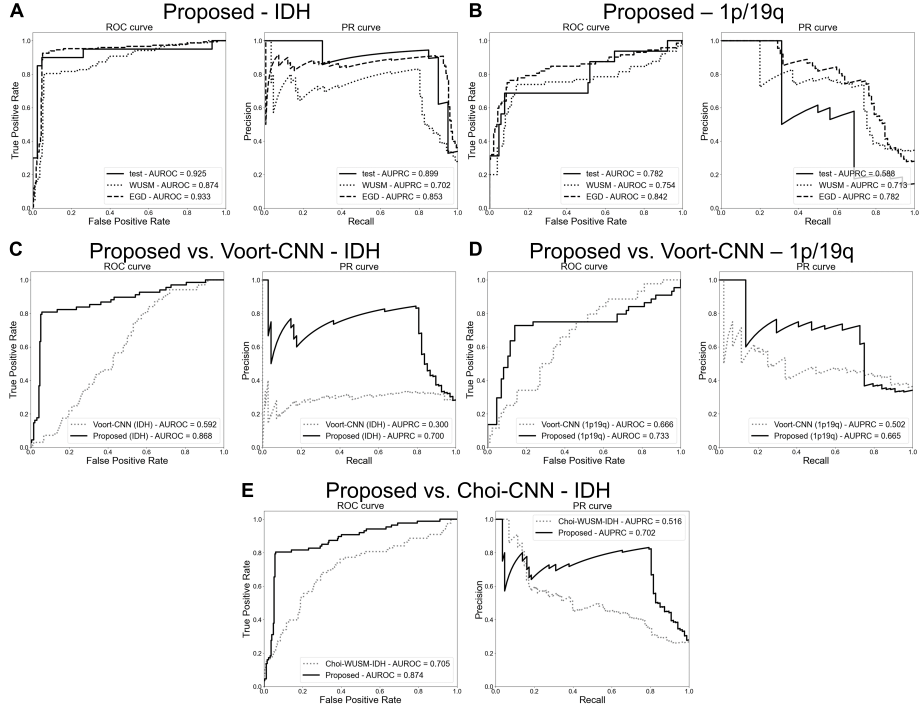}
\caption{The proposed model achieved better classification performance in both IDH status ($P < 0.05$ for precision, recall and AUROC) and 1p/19q status ($P < 0.05$ for recall) prediction tasks than the baseline methods. ROC and PR curves are reported for (A) the proposed model (CNN+age) for IDH status prediction; (B) the proposed model (CNN+loc) for 1p/19q status prediction; (C) the comparison of the proposed model performance to Voort-CNN~\cite{van2022combined}  for IDH prediction; (D) the comparison of the proposed model performance to Voort-CNN~\cite{van2022combined} for 1p/19q prediction; and (E) the comparison of the proposed model performance to Choi-CNN~\cite{choi2021fully} for IDH prediction. ROC: Receiver Operating Characteristics, PR: Precision-Recall.} 
\label{fig3}
\end{figure}

Our ablation studies (see Supplementary data~\ref*{supp_results_ablation_IDH}) determined the 2.5D CNN+age model to be the best-performing configuration for IDH status classification (Table~\ref*{table2}, Figure~\ref*{fig3}A, Supplementary Figure~\ref*{supp_fig7}A). This model yielded high accuracies on the internal (93.5\%), WUSM (90.4\%), and EGD (94.1\%) test sets. Compared to the internal test set, the model exhibited a 0.11 drop in precision (0.831), and very minor drops in recall (0.793) and AUROC values (0.874, 95\% CI: 0.826-0.917) on the WUSM test set. For the EGD set, it yielded a similar precision (0.908), and minor improvements in recall (0.926) and AUROC (0.933, 95\% CI: 0.0.902-0.960) compared to the internal test set. Overall, the model showed good generalization on both external test sets, with the performance being slightly better in EGD compared to WUSM. 

Compared to the Voort-CNN and Choi-CNN baselines, the proposed model performed significantly better. It had a higher precision (0.391 increase; $P < 0.001$), recall (0.22 increase; $P = 0.004$), and AUROC (0.113 increase; $P = 0.002$) than Voort-CNN (Figure~\ref*{fig3}C, Supplementary Table~\ref*{supp_table1}, Supplementary Figure~\ref*{supp_fig7}C). Similar improvements were obtained compared to Choi-CNN (Figure~\ref*{fig3}E, Supplementary Table~\ref*{supp_table1}, Supplementary Figure~\ref*{supp_fig7}E). Specifically, the proposed model had a higher precision (0.393 increase; $P < 0.001$), recall (0.149 increase; $P = 0.028$), and AUROC (0.17 increase; $P < 0.001$).

\subsubsection{Failure analysis and correlation with overall survival}
We identified the following main sources of error. First, given that the classification predictions are contingent upon successful tumor detections, we observed that the model failed to make any molecular status classification due to undetected tumors for a small percentage of cases (8\% and 3.8\% of all misclassified cases in WUSM and EGD sets respectively; columns marked with ‘BG’, or background, in Supplementary Figure~\ref*{supp_fig7}A). Second, classifications were sometimes affected by poor off-plane resolution, specifically in the EGD test set (36\% of the IDH-mut cases misclassified as IDH-wt) (Supplementary data~\ref*{supp_results_miscl_IDH}).

\begin{figure}[htbp]
\centering
\includegraphics[width=\textwidth]{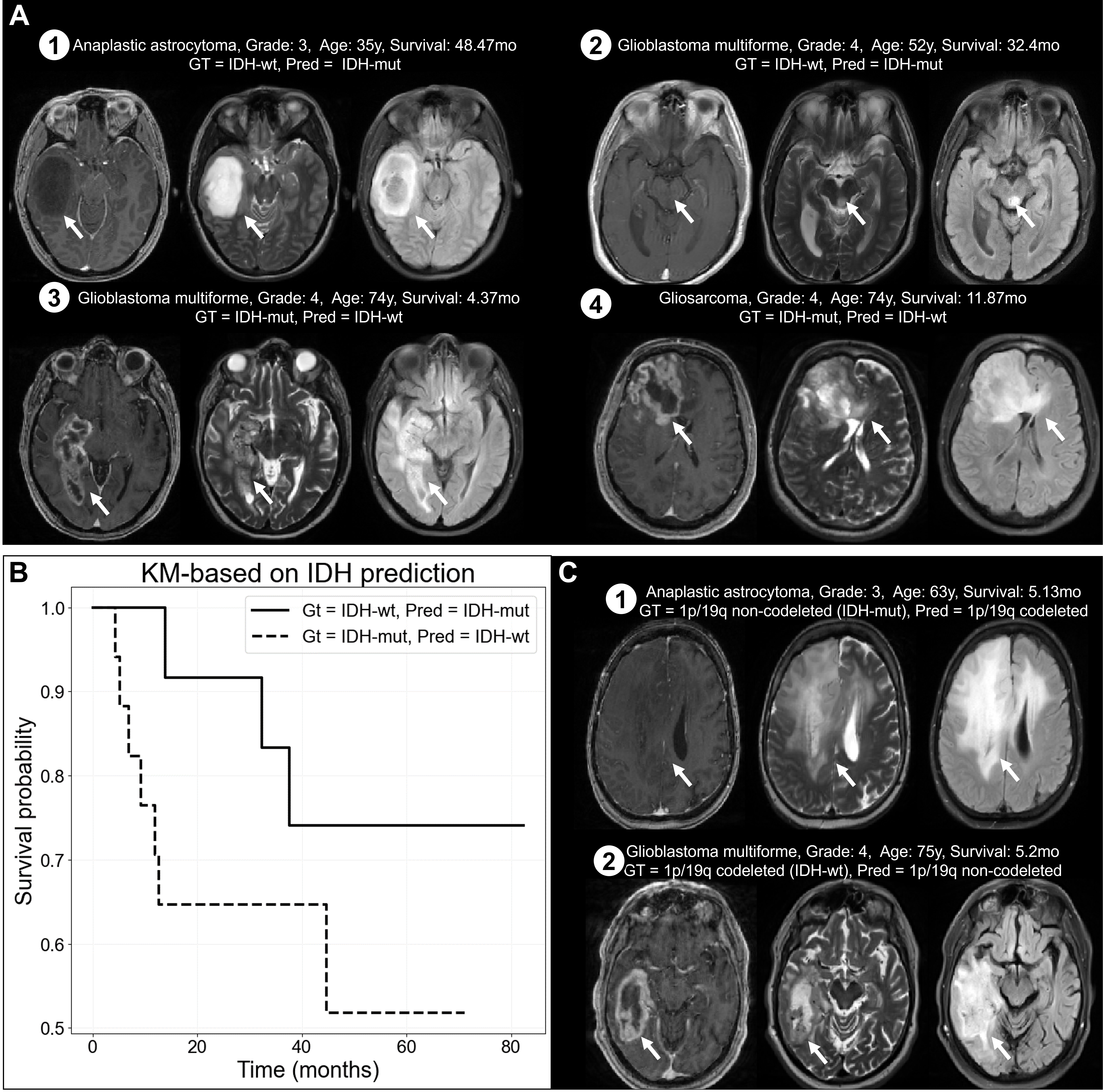}
\caption{Case study of model misclassifications in IDH and 1p/19q prediction tasks. (A) T1c, T2 and FLAIR axial slices of four exemplary tumor cases which were misclassified in IDH prediction, (B) Kaplan-Meier survival curves characterizing the overall survival for predicted IDH-mut and predicted IDH-wt groups from misclassified cases of WUSM test set in the IDH prediction task, and (C) T1c, T2 and FLAIR axial slices of two exemplary tumor cases which were misclassified in 1p/19q prediction.} 
\label{fig4}
\end{figure}

\begin{figure}[htbp]
\centering
\includegraphics[width=\textwidth]{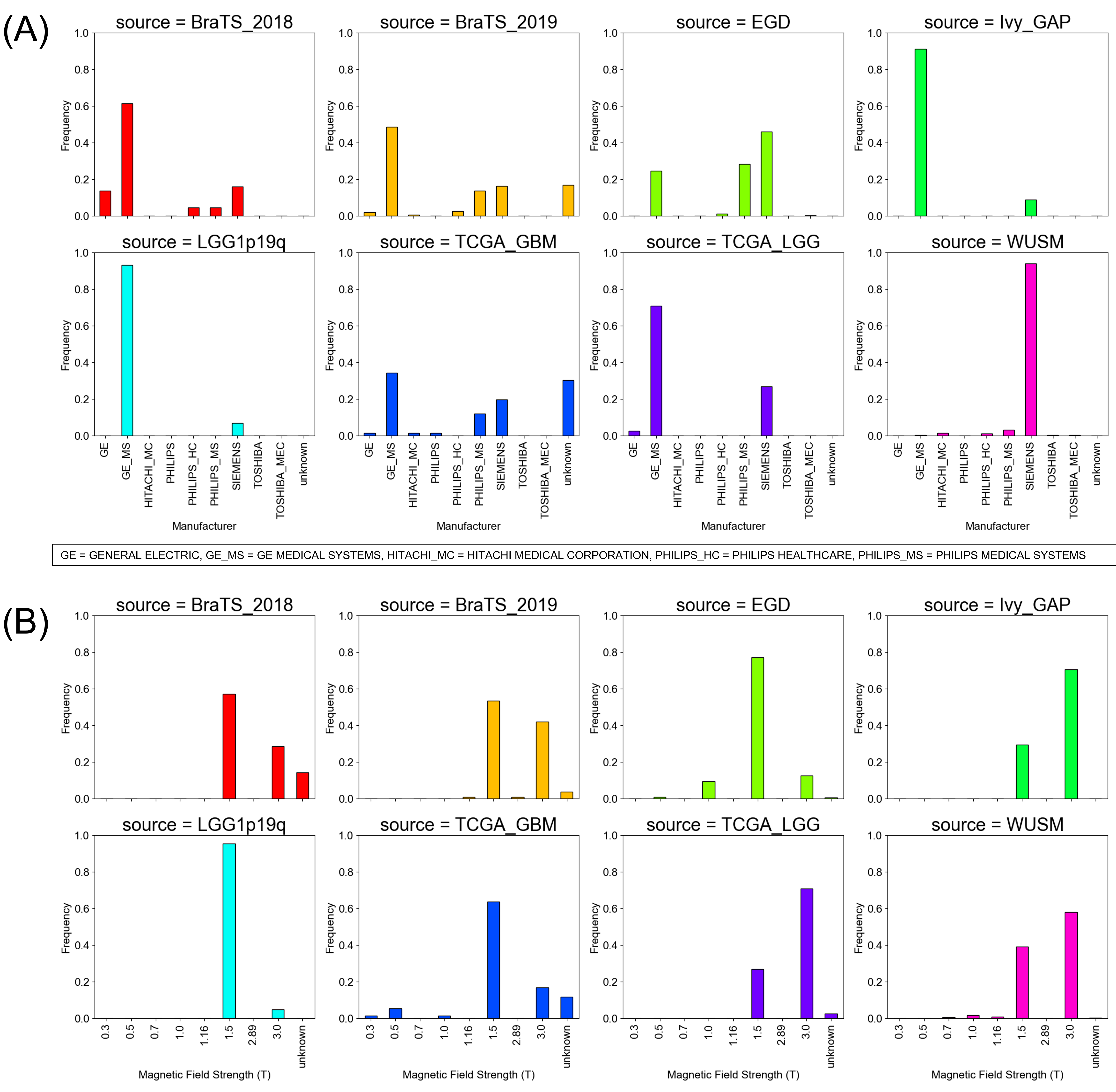}
\caption{(A) Manufacturer, and (B) Magnetic field strength of scans acquired from BraTS 2018, BraTS 2019, EGD, Ivy GAP, LGG 1p/19q, TCGA-GBM, TCGA-LGG, and WUSM datasets. BraTS = Brain Tumor Segmentation challenge, EGD = Erasmus Glioma Database, Ivy GAP = Ivy Glioblastoma Atlas Project, TCGA = The Cancer Genome Atlas, WUSM = Washington University School of Medicine.}
\label{supp_fig1}
\end{figure}

\begin{figure}[htbp]
\centering
\includegraphics[width=\textwidth]{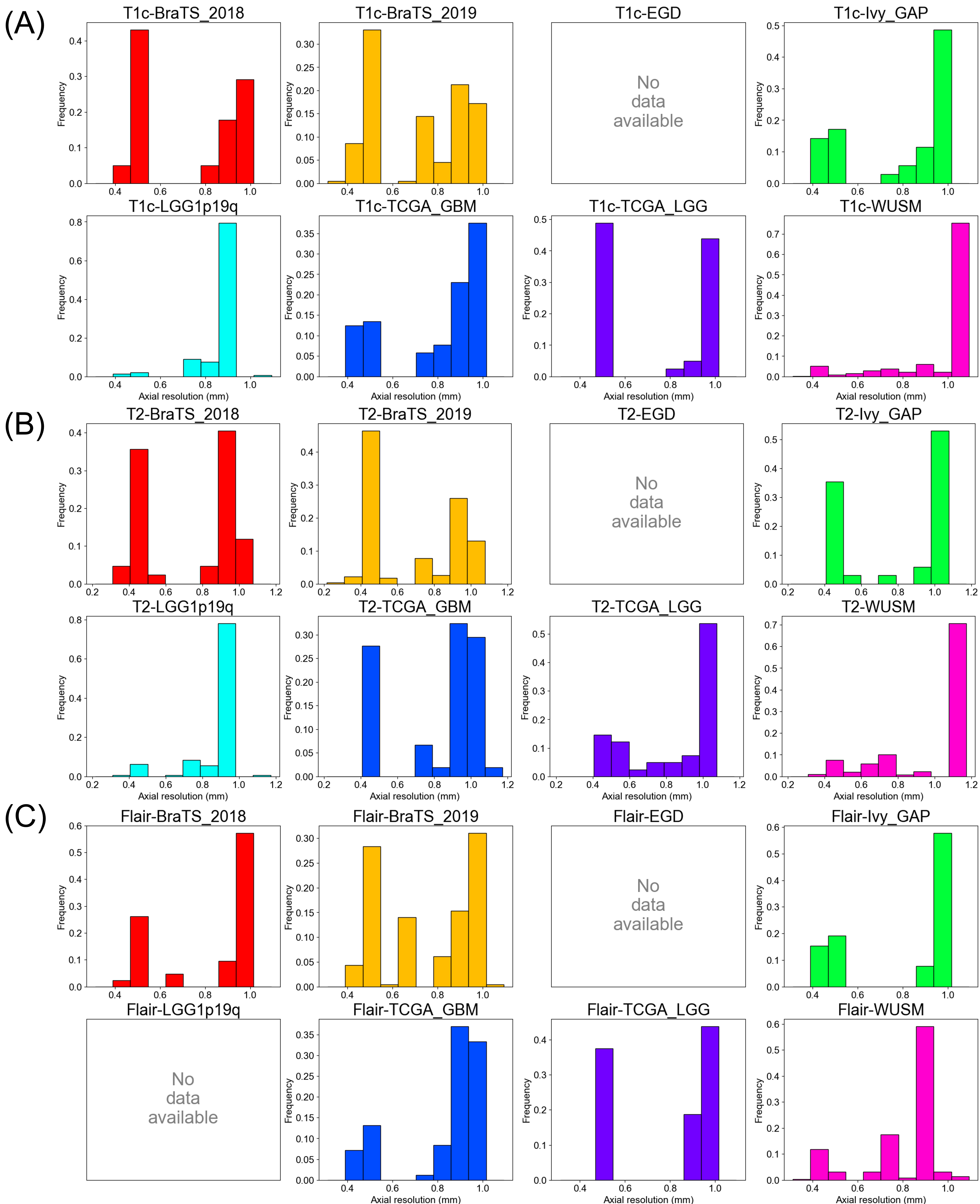}
\caption{Axial plane resolution of (A) T1c, (B) T2, and (C) FLAIR scans acquired from BraTS 2018, BraTS 2019, EGD, Ivy GAP, LGG 1p/19q, TCGA-GBM, TCGA-LGG, and WUSM datasets. T1c = post-contrast T1-weighted sequence, T2 = T2-weighted sequence, FLAIR = Fluid Attenuated Inversion Recovery sequence, BraTS = Brain Tumor Segmentation challenge, EGD = Erasmus Glioma Database, Ivy GAP = Ivy Glioblastoma Atlas Project, TCGA = The Cancer Genome Atlas, WUSM = Washington University School of Medicine.} 
\label{supp_fig2}
\end{figure}

\begin{figure}[htbp]
\centering
\includegraphics[width=\textwidth]{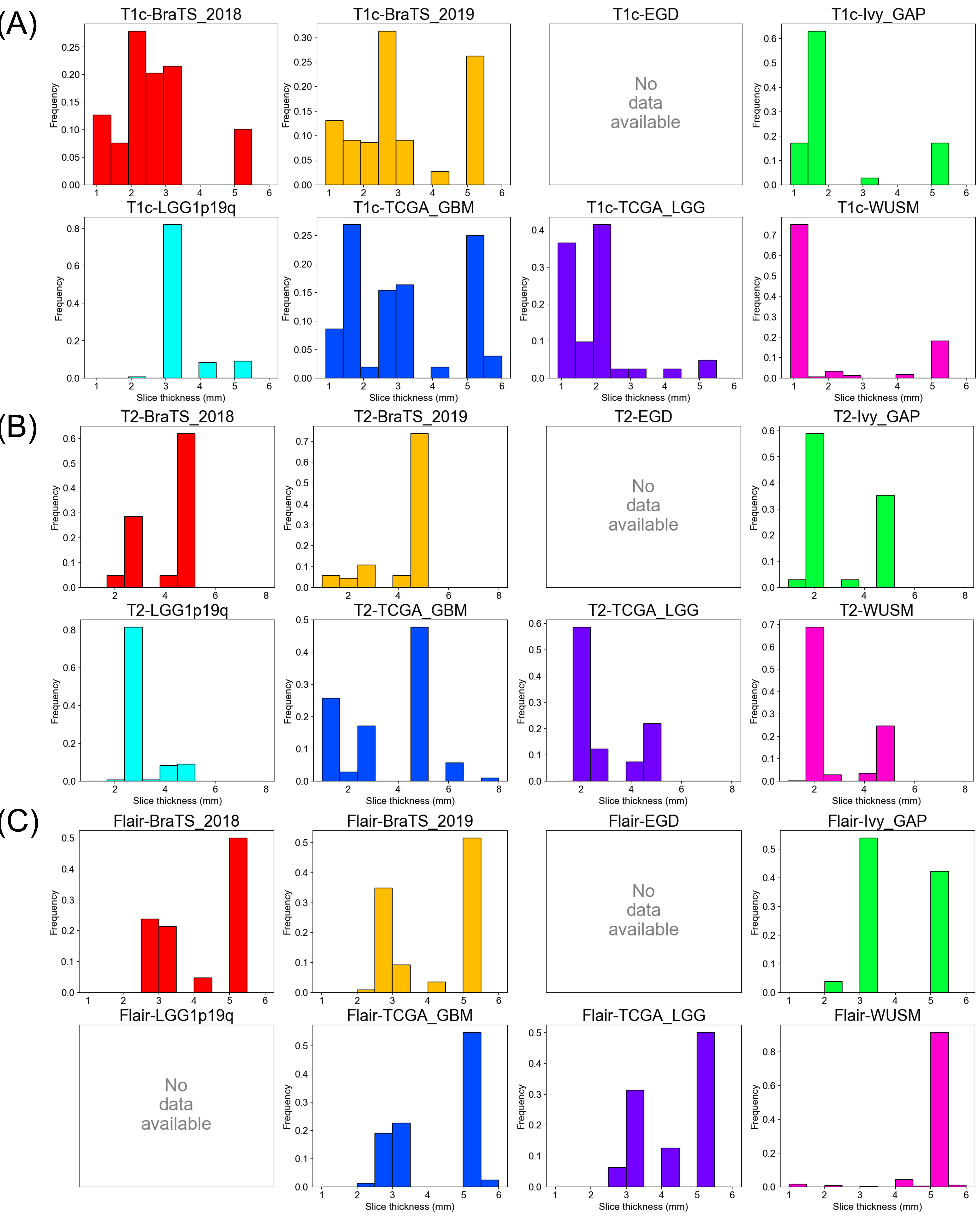}
\caption{Slice thickness of (A) T1c, (B) T2, and (C) FLAIR scans acquired from BraTS 2018, BraTS 2019, EGD, Ivy GAP, LGG 1p/19q, TCGA-GBM, TCGA-LGG, and WUSM datasets. T1c = post-contrast T1-weighted sequence, T2 = T2-weighted sequence, FLAIR = Fluid Attenuated Inversion Recovery sequence, BraTS = Brain Tumor Segmentation challenge, EGD = Erasmus Glioma Database, Ivy GAP = Ivy Glioblastoma Atlas Project, TCGA = The Cancer Genome Atlas, WUSM = Washington University School of Medicine.} 
\label{supp_fig3}
\end{figure}

\begin{figure}[htbp]
\centering
\includegraphics[width=\textwidth]{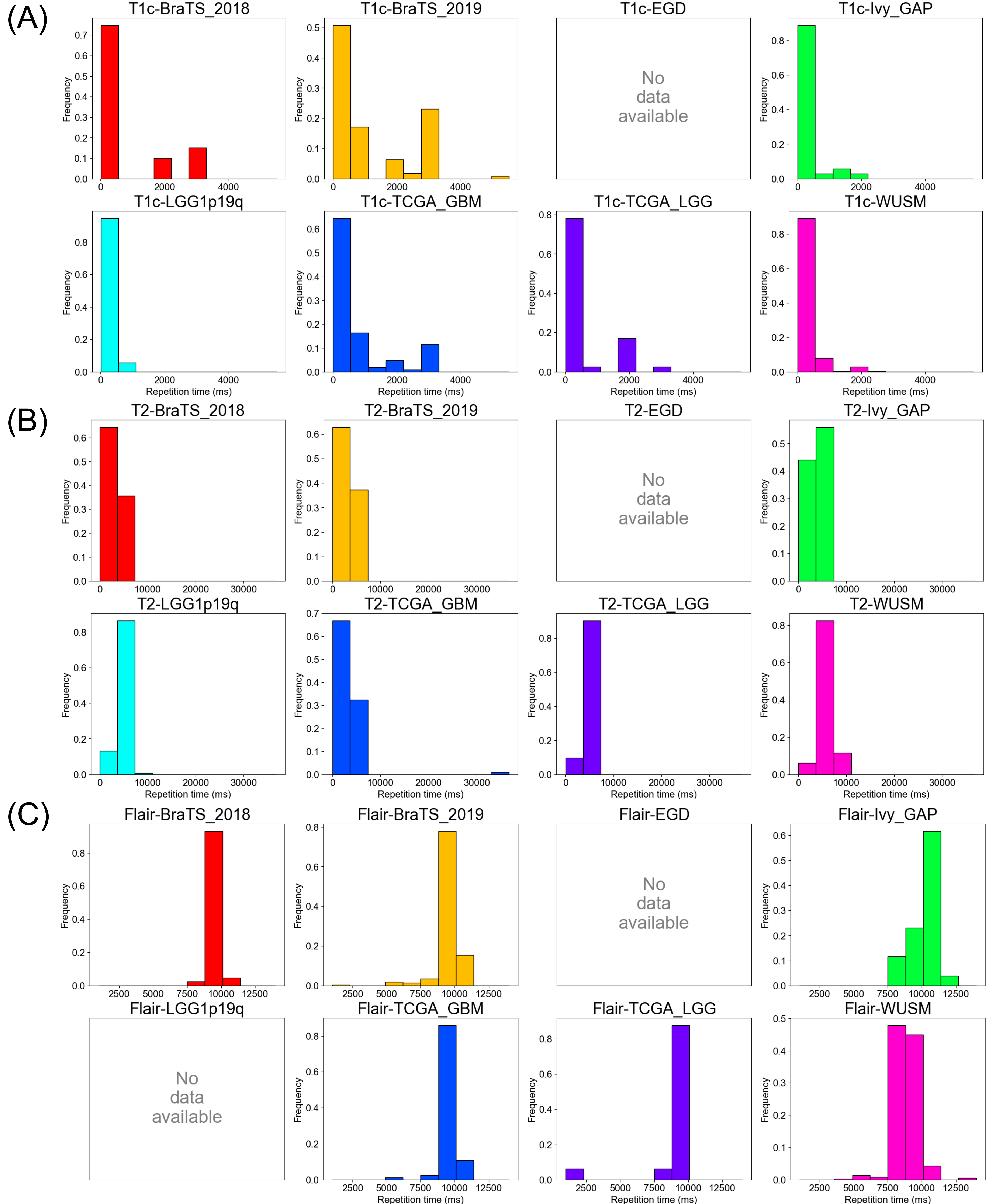}
\caption{Time of repetition (TR) of (A) T1c, (B) T2, and (C) FLAIR scans acquired from BraTS 2018, BraTS 2019, EGD, Ivy GAP, LGG 1p/19q, TCGA-GBM, TCGA-LGG, and WUSM datasets. T1c = post-contrast T1-weighted sequence, T2 = T2-weighted sequence, FLAIR = Fluid Attenuated Inversion Recovery sequence, BraTS = Brain Tumor Segmentation challenge, EGD = Erasmus Glioma Database, Ivy GAP = Ivy Glioblastoma Atlas Project, TCGA = The Cancer Genome Atlas, WUSM = Washington University School of Medicine.} 
\label{supp_fig4}
\end{figure}

\begin{figure}[htbp]
\centering
\includegraphics[width=\textwidth]{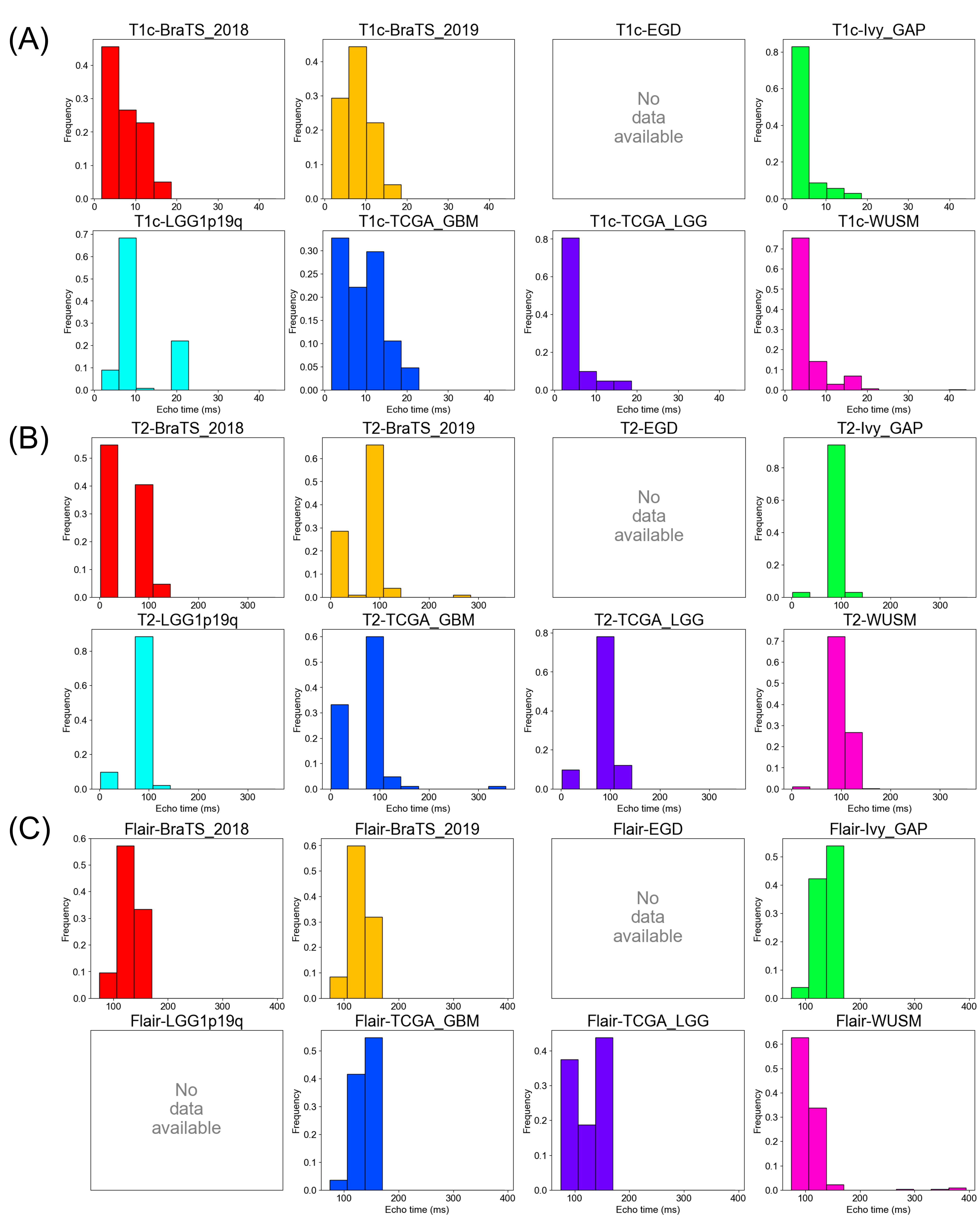}
\caption{Time of echo (TE) of (A) T1c, (B) T2, and (C) FLAIR scans acquired from BraTS 2018, BraTS 2019, EGD, Ivy GAP, LGG 1p/19q, TCGA-GBM, TCGA-LGG, and WUSM datasets. T1c = post-contrast T1-weighted sequence, T2 = T2-weighted sequence, FLAIR = Fluid Attenuated Inversion Recovery sequence, BraTS = Brain Tumor Segmentation challenge, EGD = Erasmus Glioma Database, Ivy GAP = Ivy Glioblastoma Atlas Project, TCGA = The Cancer Genome Atlas, WUSM = Washington University School of Medicine.} 
\label{supp_fig5}
\end{figure}

\begin{figure}[htbp]
\centering
\includegraphics[width=\textwidth]{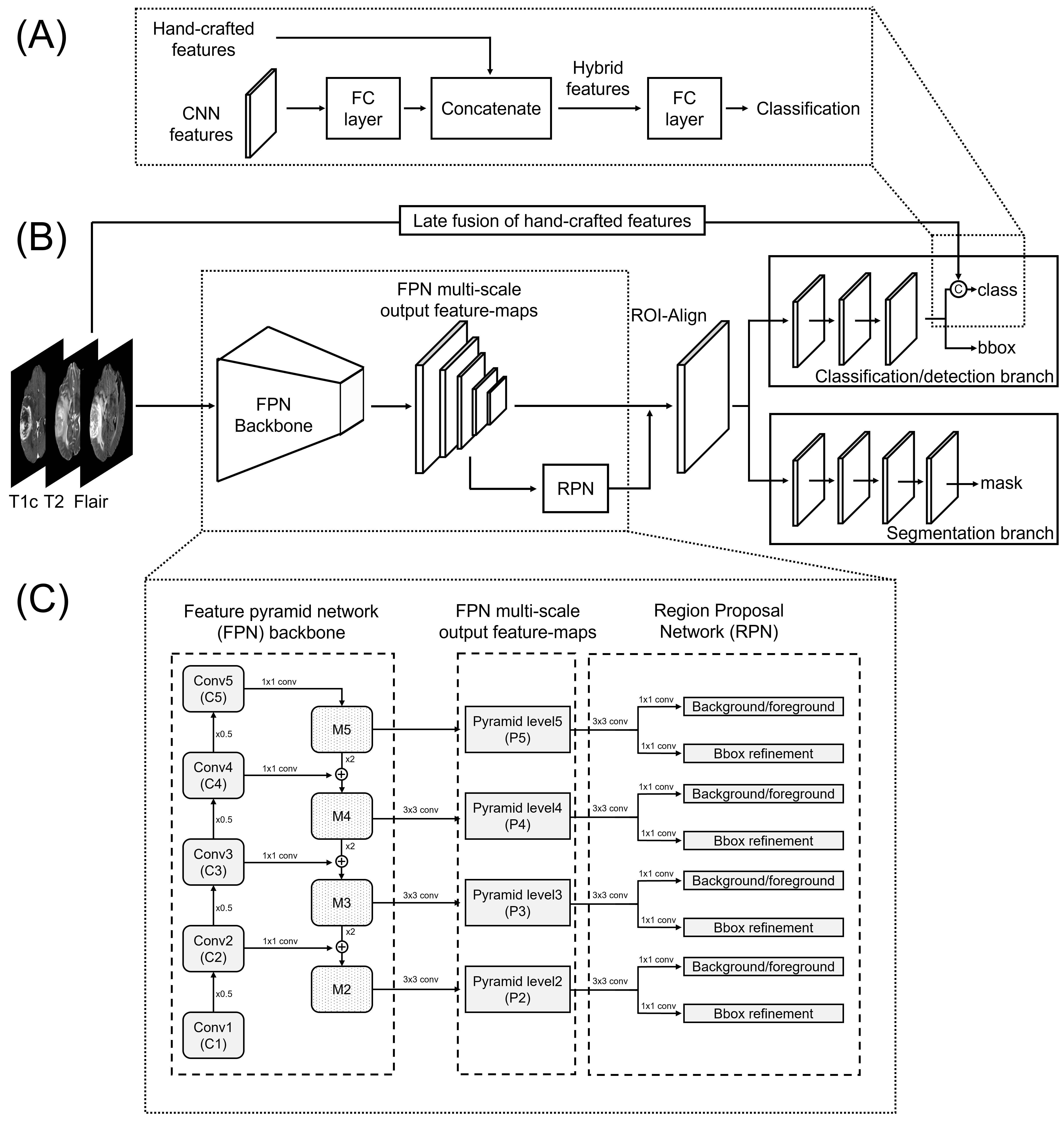}
\caption{(A) The late-fusion scheme for integrating prior knowledge features into CNN, (B) the overall hybrid Mask-RCNN architecture, and (C) the detailed architecture of the FPN and RPN. CNN: Convolutional Neural Network, FPN: Feature Pyramid Network, RPN: Region Proposal Network.} 
\label{supp_fig6}
\end{figure}

\begin{figure}[htbp]
\centering
\includegraphics[width=\textwidth]{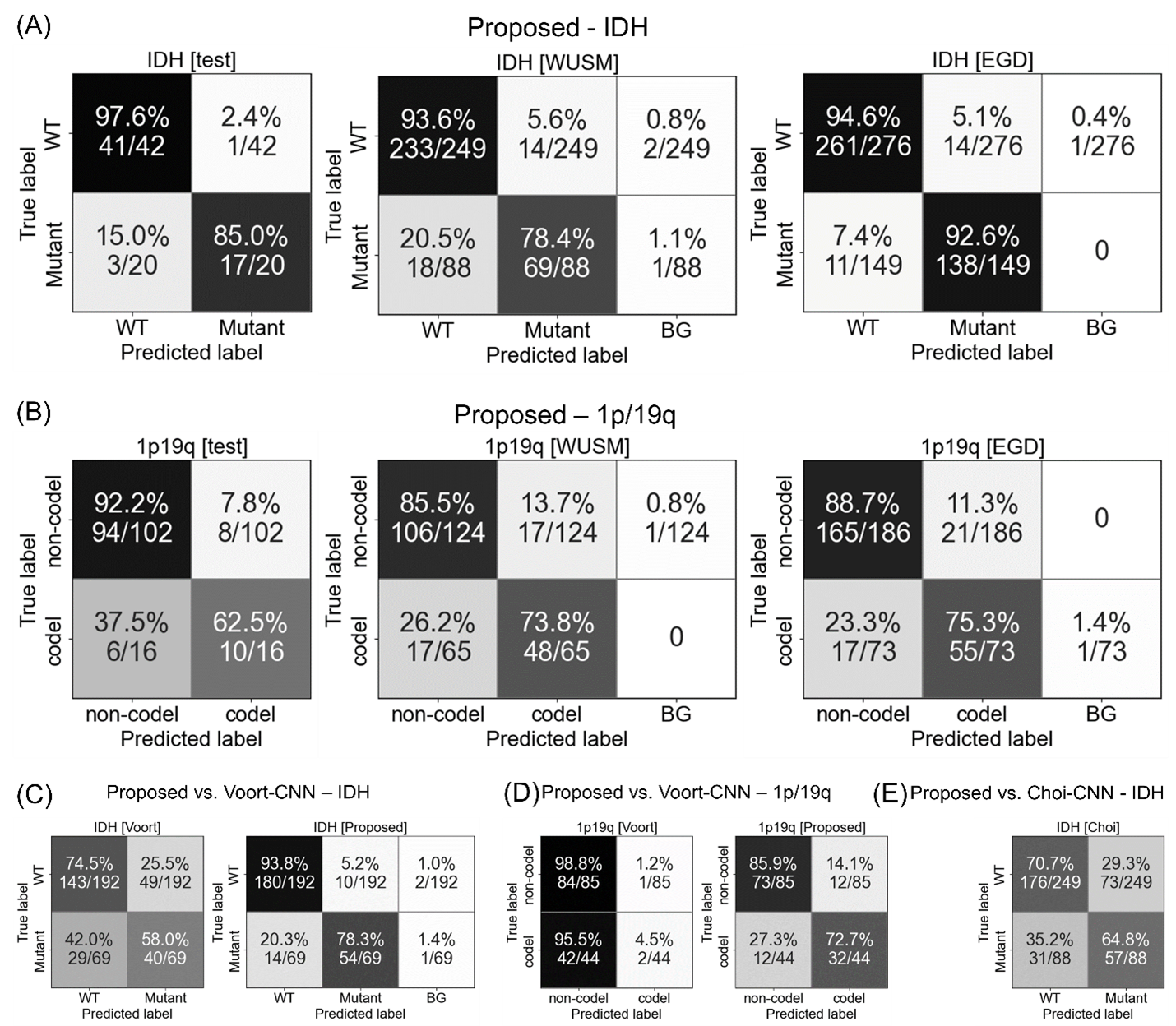}
\caption{Confusion matrices for (A) model performance for IDH status prediction, (B) model performance for 1p/19q prediction, (C) comparison of model performance to Voort-CNN~\cite{van2022combined} for IDH prediction, (D) comparison of model performance to Voort-CNN~\cite{van2022combined} for 1p/19q prediction, and (E) comparison of model performance to Choi-CNN~\cite{choi2021fully} for IDH prediction.} 
\label{supp_fig7}
\end{figure}

\begin{figure}[htbp]
\centering
\includegraphics[width=\textwidth]{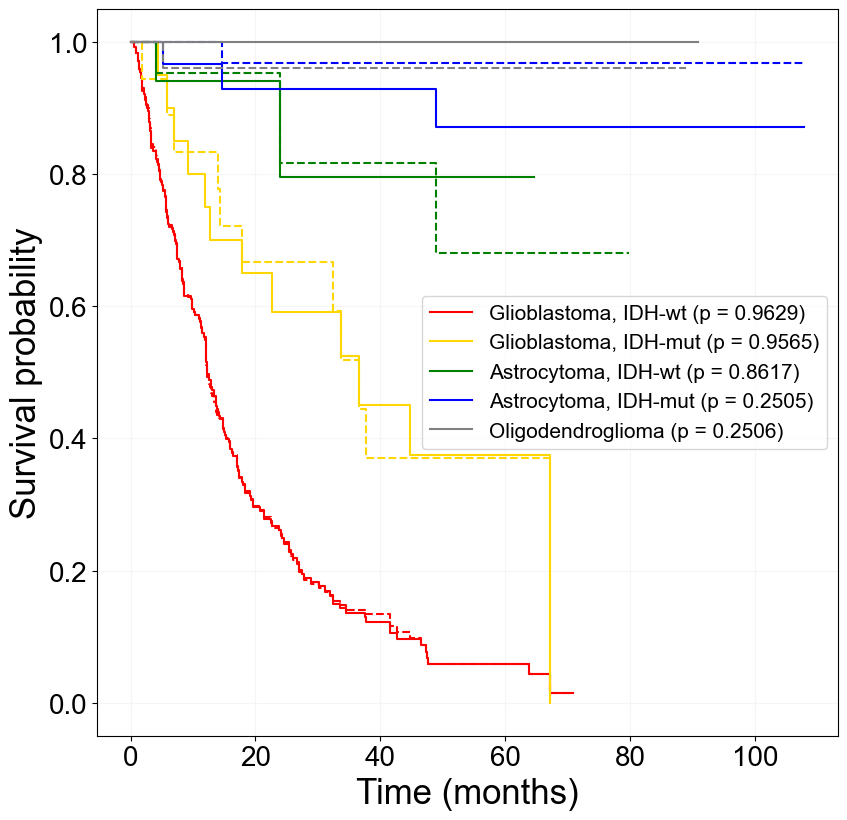}
\caption{Kaplan-Meier survival curves characterizing the OS for WHO 2016 glioma subtypes. For each subtype, the solid and dashed lines characterize the OS based on ground truth and predicted molecular status respectively. The p-values in the parentheses for each subtype (all $P > .05$) demonstrate that there is no statistically significant difference in the OS values between each ground truth and predicted pair.} 
\label{supp_fig8}
\end{figure}

Comparison of OS demonstrated that there was a high alignment between the OS of patients based on ground truth WHO 2016 subtypes and predicted subtypes (Supplementary Figure~\ref*{supp_fig8}). Analysis of the misclassifications showed that for most of the misclassified cases, the predicted IDH status had a better concordance than the IDH ground-truth label with tumor phenotype, patient age at diagnosis, and OS (Figure~\ref*{fig4}A, Supplementary data~\ref*{supp_results_miscl_IDH}). Overall, the group predicted as IDH-mut had a higher median OS than the one predicted as IDH-wt (47.6 months vs. 16.94 months) (Figure~\ref*{fig4}B).

\begin{table}[htbp]
\scalebox{0.8}{
\begin{threeparttable}
\caption{Performance of proposed model and comparison with Voort-CNN and Choi-CNN for prediction of IDH mutation and 1p/19q codeletion status.}
\label{table2}
\begin{tabular}{l|c|c|c|c|c|c}
\toprule
                      & Accuracy  & Precision & Recall    & F1-score  & AUROC (95\% CI)      & AUPRC (95\% CI)      \\ \midrule
IDH                   & \textbf{} & \textbf{} & \textbf{} & \textbf{} & \textbf{}            & \textbf{}            \\ 
~~~~~test                  & 0.935     & 0.944     & 0.85      & 0.895     & 0.925 (0.809, 1.000) & 0.899 (0.740, 1.000) \\ 
~~~~~WUSM                  & 0.904     & 0.831     & 0.793     & 0.812     & 0.874 (0.826, 0.917) & 0.702 (0.600, 0.812) \\ 
~~~~~EGD                   & 0.941     & 0.908     & 0.926     & 0.917     & 0.933 (0.902, 0.960) & 0.853 (0.780, 0.918) \\ \hline
1p/19q                &           &           &           &           &                      &                      \\ 
~~~~~test                  & 0.881     & 0.579     & 0.687     & 0.628     & 0.782 (0.627, 0.916) & 0.588 (0.354, 0.810) \\ 
~~~~~WUSM                  & 0.819     & 0.738     & 0.738     & 0.738     & 0.754 (0.666, 0.840) & 0.713 (0.611, 0.813) \\ 
~~~~~EGD                   & 0.853     & 0.724     & 0.764     & 0.743     & 0.842 (0.776, 0.904) & 0.782 (0.697, 0.860) \\ \hline
Comparison with Voort &           &           &           &           &                      &                      \\ 
~~~~~Voort-CNN   (IDH)     & 0.701     & 0.449     & 0.58      & 0.506     & 0.592 (0.522, 0.668) & 0.3 (0.237, 0.402)   \\ 
~~~~~Proposed   (IDH)      & 0.907     & 0.844     & 0.794     & 0.818     & 0.868 (0.810, 0.921) & 0.7 (0.585, 0.826)   \\ 
~~~~~Voort-CNN   (1p/19q)  & 0.667     & 0.667     & 0.045     & 0.085     & 0.666 (0.568, 0.759) & 0.502 (0.377, 0.654) \\ 
~~~~~Proposed   (1p/19q)   & 0.814     & 0.727     & 0.727     & 0.727     & 0.733 (0.630, 0.834) & 0.665 (0.539, 0.807) \\ \hline
Comparison with Choi  &           &           &           &           &                      &                      \\ 
~~~~~Choi-CNN   (IDH)      & 0.691     & 0.438     & 0.648     & 0.523     & 0.705 (0.636, 0.767) & 0.516 (0.418, 0.612) \\ \bottomrule
\end{tabular}
\begin{tablenotes}
      \small
      \item Abbreviations: IDH, Isocitrate dehydrogenase; WUSM, Washington University School of Medicine; EGD, Erasmus Glioma Database; AUROC, area under the receiver operating characteristic curve; AUPRC, area under the precision-recall curve; CI, confidence interval.
    \end{tablenotes}
  \end{threeparttable}
  }
\end{table}

\subsection{1p/19q codeletion status prediction}
\subsubsection{Classification performance}
Our ablation studies (see Supplementary data~\ref*{supp_results_ablation_1p19q}) determined the 2.5D CNN+loc model to be the best-performing configuration for 1p/19q codeletion status classification (Table~\ref*{table2}, Figure~\ref*{fig3}B, Supplementary Figure~\ref*{supp_fig7}B). This model achieved high accuracy values on the internal (88.1\%) and EGD (85.3\%) test sets, with a minor drop on the WUSM (81.9\%) set. Precision and recall metrics in the internal test set were affected by a small percentage of false-positive cases (7.8\%, 8 of 102) due to the high class imbalance therein (13.5\% codeleted vs. 86.5\% non-codeleted). Compared to the internal test set (57.9\%), the precision was much higher on the WUSM (73.8\%) and EGD (72.4\%) sets. Similarly, model recall was much higher on the WUSM (73.8\%) and EGD (76.4\%) sets compared to the internal test set (62.5\%). This led to a much higher AUPRC for the WUSM (0.713, 95\% CI: 0.611-0.813) and EGD (0.782, 95\% CI: 0.697-0.860) sets compared to the internal (0.588, 95\% CI: 0.354-0.810) test set. However, the model achieved similar AUROC values for all three datasets (0.782, 95\% CI: 0.627-0.916 for the internal test set; 0.754, 95\% CI: 0.666-0.840 for the WUSM test set; and 0.842, 95\% CI: 0.776-0.904 for the EGD test set). This disparity in AUROC and AUPRC can be explained by the severe class-imbalance in the internal test data (see Supplementary data~\ref*{supp_results_metric_details_1p19q}). Overall, the model performed better on both external test sets compared to the internal test set, with the performance in EGD being slightly better than the one in WUSM.

Compared to the Voort-CNN (Figure~\ref*{fig3}D, Supplementary Table~\ref*{supp_table2}, Supplementary Figure~\ref*{supp_fig7}D), the proposed model showed minor improvement in terms of precision (0.06 increase; $P = 0.827$) and AUROC (0.06 increase; $P = 0.371$) and a significant improvement in recall (0.682 increase; $P < 0.001$).

\subsubsection{Failure analysis and correlation with overall survival}
We observed that the model failed to make a tumor detection, and hence any subsequent molecular status classification, for a small percentage of cases (2.8\% and 2.6\% of all misclassified cases in WUSM and EGD sets respectively; columns marked with ‘BG’ in Supplementary Figure~\ref*{supp_fig7}B). 

For the other misclassifications, no discernible patterns could be identified. However, in the 13.7\% (17/124) 1p/19q non-codeleted cases in WUSM test set that were misclassified as codeleted, we found a predominance of IDH-mut cases compared to IDH-wt (11/17 IDH-mut, 3/17 IDH-wt, 3/17 IDH status unknown). Additionally, certain cases showed typical features of 1p/19q codeletion like frontal location, heterogeneous texture, and cortical infiltration (Figure~\ref*{fig4}C – case1). Of the 26.2\% (17 of 65) 1p/19q codeleted cases in WUSM test set misclassified as non-codeleted, we found 5 grade-IV glioblastoma cases that were IDH-wt and had low survival (median OS 5.2 months, range 0.1-13.4 months) (Figure~\ref*{fig4}C – case2). This genetic-histologic combination is more consistent with 1p/19q non-codeletion.

\section{Discussion}
We developed a deep learning model for classification of IDH mutation and 1p/19q codeletion status that combines prior clinical knowledge and imaging features through a hybrid CNN architecture. To the best of our knowledge, the proposed method has been validated on the largest dataset till date, obtained from one clinical and seven public sources. The model achieved high accuracy on this heterogeneous dataset and showed excellent generalization capabilities on unseen testing data. The code and pre-trained models of this work will be made available for community use.

Previous studies have explored the association between tumor phenotype and molecular status. Qualitative analyses have examined visual signatures from MR sequences according to the Visually AcceSAble Rembrandt Images (VASARI) guidelines or the T2-FLAIR mismatch signature~\cite{patel2017t2}. Quantitative analysis have investigated combining radiomic features and ML~\cite{van2021accuracy,bhandari2021noninvasive}. Though ML models have been shown to perform better than visual analysis~\cite{zhou2017mri}, they still require manual intervention due to extensive feature engineering and selection. Hence, they often suffer from lack of reproducibility on new datasets. In contrast to both visual and ML approaches, our CNN-based workflow is completely end-to-end, does not require any manual intervention, automatically learns hierarchical features, integrates readily available clinical information and shows great generalization on external datasets. 
Previous studies have also explored CNN-based approaches for predicting the molecular landscape of gliomas. Our study improves previous work in several ways. First, unlike previous studies~\cite{matsui2020prediction,pasquini2021deep}, which suffered from small sample size or lack of external validation, the generalizability of our model was validated on the largest external dataset till date, comprising 968 patient cases from 11 institutions. Second, previous methods often required either a previously segmented tumor mask~\cite{chang2018deep,chang2018residual,decuyper2021automated} or a manually extracted bounding box~\cite{pasquini2021deep} around the tumor for classification. In contrast, our model simultaneously detects and segments the tumor and classifies its molecular status. This multi-task approach obviates the requirement of any prior tumor segmentation and enables the model to learn context from multiple related tasks. Third, our model is agnostic to glioma grade and thus moves substantially beyond prior efforts~\cite{matsui2020prediction,li2017deep} that focused on specific grades of glioma. This facilitates the clinical translation of our model as the tumor grade is unknown in the clinical pre-operative setting. Fourth, objective comparison between different methods is hindered by usage of different datasets and performance metrics. To address this, we performed head-to-head comparisons between our method and two recent approaches~\cite{choi2021fully,van2022combined}. 

Specifically, we used an independent dataset to explore methodological, computational and performance advantages of the proposed method compared to the works of Voort et al.~\cite{van2022combined} and Choi et al.~\cite{choi2021fully}. Our model achieved significantly better overall performance compared to the multi-task CNN method by Voort et al.~\cite{van2022combined}. As our ablation studies also demonstrated, this is due to our hybrid model’s ability to jointly learn from images as well as knowledge distilled from clinical records and neuroanatomical information. A hybrid approach was also proposed by Choi et al.~\cite{choi2021fully}. However, their model combined radiomic features with a 2D CNN to predict IDH status, thus not providing a full classification of the gliomas, and required a separate CNN for tumor segmentation. In comparison with Choi-CNN, the proposed model yielded significantly higher overall performance. This improvement can be attributed to the usage of 2.5D models, which captures the 3D spatial context of the brain, while being computationally efficient. This also supported by our ablation studies that showed that the 2.5D model performed significantly better than the 2D planar models for both prediction tasks.  

In an overall comparison between the IDH and 1p/19q classification performances, we found that the models generally yielded better results for IDH. This is in line with a recent review of radiogenomic studies~\cite{van2021accuracy} that observed a significantly poorer 1p/19q classification performance compared to other molecular subtypes. For IDH status classification, multiple studies~\cite{chang2018deep,zhou2017mri,eckel2015glioma} have associated IDH-wt gliomas with thick, irregular, and poorly marginated enhancement on T1c scan. In contrast, IDH-mut gliomas have been associated with minimal or no enhancement on T1c, and well-defined tumor margins. There is also evidence~\cite{zhou2017mri,eckel2015glioma,yan2009idh1} of lower age of diagnosis in patients with IDH-mut gliomas compared to IDH-wt. In our study, analysis of cases with misclassified IDH status showed that this existing knowledge of age, tumor phenotype and OS trends was better aligned with the predicted class than with the ground truth. This alludes to possible errors in the histopathological assessment of the tumor molecular status originating from variability in cut-off values used to determine IDH status in immunohistochemistry (IHC) evaluations~\cite{lee2013idh1}, heterogeneity of staining in IHC leading to partial uptakes~\cite{agarwal2013comparative}, or heterogeneity in samples where only a fraction of tumor cells have IDH1-R132H expression~\cite{preusser2011value}. For 1p/19q status classification, several 1p/19q codeleted cases that were misclassified as non-codeleted were in fact glioblastomas with low OS. This suggests possible histopathological false-positive assessment for these cases caused by a partial 1p/19q codeletion~\cite{ball2020frequency} being misclassified by the fluorescence in situ hybridization (FISH) technique due to its inability to distinguish partial from whole-arm deletions. Partial deletions, specifically interstitial and terminal 1p deletions, have been suggested to be particularly common in glioblastomas and are known to confound the FISH assay~\cite{horbinski2011gone}. Besides these possible errors, the histopathological assessment relies on invasively and locally obtained tissue samples. In contrast, the proposed work offers several advantages. First, our workflow can perform a non-invasive pre-operative determination of molecular status that can inform clinical decision-making and lead to a better overall survival~\cite{patel2019extent,wijnenga2018impact}. Second, the proposed model can enable fast, cost-effective tumor characterization that can be particularly useful in low-resource settings. Third, it can be especially useful for patients with certain risk factors for biopsy (e.g., due to old age or other neurological conditions) or tumors which are difficult to operate on (e.g., due to location in eloquent brain). Fourth, besides pre-operative treatment planning, this model can be used for repeated evaluation of the molecular status, thus allowing longitudinal characterization of tumor without any associated invasive interventions. Overall, in this emerging era of precision diagnostics, this workflow can drive personalized treatment planning by streamlining molecular characterization of gliomas.

There are certain limitations in this study that merit discussion. First, studies have shown the importance of tumor blood flow information from perfusion imaging~\cite{yamashita2016mr}, or detection of 2-HG within tumor through MR-spectroscopy~\cite{choi20122} in IDH prediction. However, in this work, we included only routine MR sequences and did not use any advanced sequences as these are often not included in clinical tumor protocol. This makes clinical translation of our model easier, while allowing us to leverage a much bigger dataset to train and validate our model. Second, we have included IDH and 1p/19q in this study as they are the two most important factors in classification of glioma according to the WHO 2016 guidelines. However, after 2016, WHO has updated the guidelines~\cite{louis20212021} to further refine this classification and emphasized the importance of other molecular markers like O6-methylguanine-methyltransferase (MGMT) methylation status, epidermal growth factor receptor (EGFR), p53 mutation, mutations in telomerase reverse transcriptase (TERT) and ATP-dependent X-linked helicase (ATRX) etc. These markers have not been included in this study and can be included in a future work.

In conclusion, we developed a CNN model that can classify IDH mutation and 1p/19q co-deletion status from pre-operative structural MR sequences. The model can be extended to predict other molecular alterations that are associated with specific phenotypical signatures on MR images. The network provides an important step towards developing an artificial intelligence–augmented neuro-oncology workflow that can pre-operatively predict tumor behavior and assist treatment planning leading to better outcomes.



\bibliographystyle{ama}
\bibliography{mybibliography}

\section*{\hfil Supplementary Data \hfil}

\beginsupplement

\section{Supplementary Methods}
\subsection{Molecular status ground truths for datasets} \label{supp_methods_gt}
Genetic and histological data were available for all public sources. For the in-house WUSM dataset, IDH status is determined using immunohistochemistry (monoclonal antibody for R132H) routinely as a first pass which is able to pick up the most common mutation (R132H) in 90\% cases. If that is negative, then samples are sent for targeted next-generation sequencing (tNGS) to pick up non-canonical variants. For determining 1p/19q status, Fluorescence in situ hybridization (FISH) was performed on paraffin-embedded tissue with locus-specific commercial probes (Abbott Molecular, Des Plaines, IL) localizing to 1p36, 1q25, 19p13 and 19q13.

\subsection{Data acquisition parameters} \label{supp_methods_dataacq}
The most common parameters along with their percentage in the corresponding datasets are as follows: magnetic field strength: 1.5T in BraTS 2018 (57\%), 1.5T in BraTS 2019 (53\%), 1.5T in EGD (77\%), 3T in Ivy GAP (71\%), 1.5T in LGG-1p/19q (95\%), 1.5T in TCGA-GBM (64\%), 3T in TCGA-LGG (71\%), 3T in WUSM (58\%); scanner manufacturer: General Electric (GE) in BraTS 2018 (61\%), GE in BraTS 2019 (49\%), Siemens in EGD (46\%), GE in Ivy GAP (91\%), GE in LGG-1p/19q (93\%), GE in TCGA-GBM (34\%), GE in TCGA-LGG (71\%), Siemens in WUSM (94\%); T1c slice thickness: 2.0–3.0 mm in BraTS 2018 (41\%), BraTS 2019 (37\%), LGG-1p/19q (82\%), TCGA-GBM (31\%), and 1.0–2.0 mm in Ivy GAP (68\%), TCGA-LGG (68\%), $\leqslant$1.0 mm in WUSM (58\%); T2 slice thickness: 4.0–5.0 mm in BraTS 2018 (62\%), BraTS 2019 (73\%), TCGA-GBM (53\%),1.0–2.0 mm in Ivy GAP (60\%), TCGA-LGG (58\%), WUSM (69\%), and 2.0–3.0 mm in LGG-1p/19q (81\%); FLAIR slice thickness: 4.0–5.0 mm in BraTS 2018 (50\%), BraTS 2019 (55\%), TCGA-GBM (55\%), TCGA-LGG (50\%), WUSM (93\%), and 2.0–3.0 mm in Ivy GAP (54\%). No acquisition parameters except scanner manufacturer and magnetic field strength were known for the EGD dataset.

\subsection{Data pre-processing and feature extraction} \label{supp_methods_preproc}
Scans from BraTS~\cite{menze2014multimodal,bakas2017advancing,bakas2018identifying} and Ivy GAP~\cite{puchalski2018anatomic} datasets were already registered to the SRI24 anatomical atlas~\cite{rohlfing2010sri24}, resampled to 1-mm3 isotropic resolution and skull-stripped. For consistency, raw data from TCGA-GBM, TCGA-LGG, LGG-1p/19q, and WUSM were first rigidly co-registered to the T1c scan on a per-subject basis and then registered to the SRI24 atlas~\cite{rohlfing2010sri24} using the Multimodal Glioma Analysis (MGA) pipeline~\cite{chakrabarty2020preprocessing}. The data from EGD had already been registered to the ICBM 2009a nonlinear MNI152 atlas~\cite{fonov2009unbiased,fonov2011unbiased}. Thus, we linearly registered this data to the SRI24 space using an affine transformation from the MNI152 to SRI24 atlas computed using MGA. This same transformation was applied on the accompanying tumor segmentations using a nearest-neighbor interpolation. Subsequently, for each patient, all the scans were skull-stripped using the Robust Brain Extraction~\cite{iglesias2011robust} tool. Image intensities within the brain were normalized to zero mean and unit variance after excluding intensities below the 5th and above the 95th percentile. Finally, all images were downsampled to 128x128x128 before being fed into the network. 

The age information was directly extracted from patient records. The tumor location features were extracted using the FSL ‘Atlasquery’ tool after registering the tumor segmentation mask to the “MNI structural atlas” (derived from more than 50 subjects' structural images)  provided with FSL. This generates as output nine probability values corresponding to the tumor being in one of the following nine brain regions: caudate, cerebellum, frontal lobe, insula, occipital lobe, parietal lobe, putamen, temporal lobe, and thalamus.

\subsection{Model architecture and hyperparameters} \label{supp_methods_model}
The networks used in this study are based on the Mask RCNN architecture~\cite{he2017mask}, built on top of a Residual Network-101 (ResNet-101)-Feature Pyramid Network (FPN) backbone~\cite{he2015delving,lin2017feature}, followed by a region proposal network (RPN), a layer for alignment of the region-of-interests (ROI-Align layer), and two network heads for classification/detection and segmentation (Supplementary Figure~\ref*{supp_fig6}). 
The hyperparameters of the model were determined using five-fold cross-validation. For this purpose, we split the training data into five folds using a stratified sampling strategy to maintain the original ratio of different image classes in every fold. Next, we trained the model on four folds of data (80\% of training data) and validated its performance on the held-out fold (20\% of the training data). This process was repeated for each fold in a round-robin fashion. Different sets of hyperparameters were chosen using a random search strategy~\cite{bergstra2012random}. The set of hyperparameters that yielded the best cross-validation results was then selected. Next, the model was trained on 100\% of the training data using this best set of hyperparameters. The resulting model was subsequently used for prediction on the internal and external test sets.

The network was trained in two stages for a total of 200 epochs. In the first stage, the network head (entire network except backbone, i.e., RPN, classifier and mask heads of the network) was trained for 75 epochs. In the subsequent second stage, the entire network was trained for 125 epochs. We used a batch-size of 4, a learning rate of 0.001, a Stochastic Gradient Descent (SGD) optimizer with a momentum of 0.9 and gradient clipping at 5.0. An L2-regularization of $1\times10^{-4}$ was used in all trainable layers except the batch-normalization layers. To prevent the network from overfitting, the data were augmented during training using mirroring along the vertical axis and random rotations with probabilities of 0.5. To address class-imbalance, we used a class-weighted softmax cross-entropy loss in the classification head of the model.

\subsection{Model training and testing strategies} \label{supp_methods_training}
Our training strategy involves feeding the proposed network with slices extracted from each subject’s T1c, T2, and FLAIR scans (only T1c and T2 in case of 1p/19q) along with the prior knowledge features. The subject’s multiclass tumor segmentation mask is used in three distinct ways. First, the whole tumor mask itself is used as ground-truth for segmentation. Second, the bounding box ground-truth for detection is deduced by calculating the smallest rectangle that encloses the mask. Third, the mask is used to inform the extraction of the 2D input slices from the 3D volumes as follows. First, the edema part of the tumor is excluded and the slice with the biggest TC area is selected (let n be the index of this slice). Subsequently, to naturally augment the training data, we utilize the information from neighboring slices, which share considerable information with the selected $n^{th}$ slice. Specifically, we select two additional sets of slices by extracting from all available modalities the $(n+2)^{th}$ slice and the $(n–2)^{th}$ slice, respectively. Finally, these three sets of slices (i.e., $(n+2)^{th}$, $n^{th}$ and $(n-2)^{th}$ sets) are used as three independent samples per subject, thus naturally increasing the training sample size by a factor of 3.

During testing, for each subject, all slices are extracted from available modality scans (T1c, T2, FLAIR for IDH; T1c, T2 for 1p/19q) from the three orthogonal planes. Next, for each 2D network, all slices of the corresponding plane, along with the prior knowledge features, are fed into the network. This way the entire 3D information is captured, which obviates the requirement of any prior tumor segmentation produced through manual or automatic methods. For each set of slices, the network performs three operations – detection of the tumor using a bounding box, segmentation of the whole tumor, and classification of the molecular status of the tumor. Upon running the network over all sets of slices, the classification result of the corresponding plane is determined using majority-voting across all slices where a tumor has been detected.
Once this process is repeated for axial, coronal, and sagittal planes, three classification outcomes are obtained for each input data. To determine the final classification result, these three predictions from the three planar models are ensembled using majority-voting. 

The code has been implemented in Python 3.6.2, using the Keras 2.2.5 and Tensorflow-gpu 1.12.0 libraries.

\subsection{Ablation studies} \label{supp_methods_ablation}
To determine the importance of prior knowledge features in improving the model performance, ablation studies were performed to compare performances between four network schemes viz. CNN without any prior knowledge feature (i.e., conventional CNN), CNN with ‘patient age at diagnosis’ feature (‘CNN+age’), CNN with ‘anatomical tumor location’ feature (‘CNN+loc’), and CNN with both of these features (‘CNN+age+loc’). The network scheme with the best performance among these four combinations was selected for the subsequent step.

Once the network scheme was determined, ablation studies were performed within the scheme to compare the proposed 2.5D model, which aggregates information from all three planes to individual models trained on axial, sagittal, coronal planes (planar models). This was done to determine if the aggregated information from three planes improves model performance. 
This whole process was repeated separately for the IDH and 1p/19q classification tasks.

\subsection{Baseline methods for comparison} \label{supp_methods_baselines}
The performance of the proposed model was compared to two baseline pre-trained models: (i) a multi-task U-net model by Voort et al.~\cite{van2022combined} (hereon referred to as “Voort-CNN”) for both IDH and 1p/19q prediction tasks, (ii) a CNN-radiomics hybrid model by Choi et al.~\cite{choi2021fully} (hereon referred to as “Choi-CNN”) for only the IDH prediction task. To ensure a fair comparison, the models were compared only on the WUSM test set because the EGD dataset had been used as training data for Voort-CNN in the original paper~\cite{van2022combined}. 

The Voort-CNN model requires pre-contrast T1-weighted (T1w) MRI scans along with T1c, T2, and FLAIR sequences for prediction. Based on this requirement, the original WUSM test sets for IDH prediction (n = 337) and 1p/19q prediction (n = 187) were filtered to smaller “WUSM-IDH-4modalities” (n = 261) and “WUSM-1p/19q-4modalities” (n = 129) subsets, respectively. Results for Voort-CNN were obtained using the trained model provided by the authors as the \verb|svdvoort/prognosais_glioma:1.0.2| docker. 
The Choi-CNN model had the same modality requirements as the proposed model and could be evaluated on the same WUSM test set (n = 337). This model had been trained for only IDH prediction, and hence the performance could not be compared for 1p/19q prediction. Results for Choi-CNN were obtained using the trained model provided by the authors as a github codebase (\url{https://github.com/yoonchoi-neuro/automated_hybrid_IDH}).

\subsection{Statistical analysis} \label{supp_methods_stat}
For AUROC and AUPRC, the 95\% confidence intervals (CI) were calculated using a 1000-sample bootstrapping method, i.e., by resampling the prediction scores 1000 times with replacement, calculating the scores for each of those 1000 instances and taking the 5th and 95th percentile values from the sample of 1000 values. The statistical analyses and the visualization of the results were performed using scikit-learn, numpy, pandas, seaborn, and lifelines libraries in Python 3.6.2, and the DTComPair and pROC packages in R, version 4.1.0 (\url{https://www.rproject.org/}). For all statistical analyses, the threshold for statistical significance was set to $P < .05$.

\subsection{Misclassifications and survival analysis} \label{supp_methods_survival}
For both IDH and 1p/19q classification tasks, we attempted to identify recurring patterns in the cases misclassified by the best-performing models. Prototypical 1p/19q codeleted gliomas have often been associated with frontal predominance~\cite{van2019predicting,shboul2020prediction} and heterogeneous texture~\cite{van2019predicting,batchala2019neuroimaging}. For IDH mutation, studies~\cite{zhou2017mri,chang2018deep,eckel2015glioma} have associated IDH-wt gliomas with irregular enhancements and ill-defined tumor margins, while IDH-mut gliomas with minimal or no enhancement and well-defined tumor margins. We have analyzed the misclassified cases in the light of these previously established phenotypical and prognostic characteristics.  We further hypothesized that the misclassified IDH-wt cases that have IDH-mut like phenotype will have better OS. We analyzed these cases by comparing the groups of misclassified cases (i.e., IDH-wt predicted as IDH-mut and vice versa) in terms of OS.

Additionally, for evaluating the validity of our model in classifying gliomas into WHO 2016 subtypes, each case in the dataset was classified according to the WHO 2016 guidelines~\cite{louis20162016}: IDH-mut, 1p/19q codeleted, grade II/III glioma as oligodendroglioma; IDH-mut, 1p/19q non-codeleted, grade II/III glioma as IDH-mut astrocytoma; IDH-wt, 1p/19q non-codeleted, grade II/III glioma as IDH-wt astrocytoma; IDH-mut, grade IV glioma as IDH-mut glioblastoma and IDH-wildtype, grade IV glioma as IDH-wt glioblastoma. In all cases, the glioma grade information was obtained from the clinical records. Based on the ground truth molecular status and predicted molecular status of these cases, the cases were placed into different groups and their OS was compared using Cox regression analysis. We hypothesized that the groups based on ground truth molecular status and the corresponding groups based on predicted molecular status will have no significant difference in terms of OS.

\section{Supplementary results}
\subsection{Dataset characteristics} \label{supp_results_dataset}
For the IDH subset, the internal ($P = 0.041$), WUSM ($P < 0.001$) and EGD ($P = 0.002$) test sets differed significantly in terms of age compared to the training data. In terms of sex, the internal test and training data had no significant difference ($P = 0.483$). However, the WUSM ($P = 0.041$) and EGD ($P = 0.026$) test sets differed significantly compared to training set, with a higher prevalence of male subjects. In terms of IDH status, the internal ($P = 0.283$) and EGD ($P = 0.176$) test sets showed similar distributions compared to the training data. However, the WUSM test set differed significantly with a much lower prevalence of IDH-mut cases compared to the training set (26.11\% vs. 40.81\%; $P < 0.001$). In terms of tumor grade, the internal ($P < 0.001$), WUSM ($P < 0.001$), and EGD ($P < 0.001$) test sets differed significantly compared to the training set. Compared to the training set, the internal test set had a much higher percentage of grade IV gliomas (74.19\% vs.46.64\%), and lower percentage of grade III gliomas (8.06\% vs.26.46\%). Similar trends were observed in the WUSM (grade III 13.95\%, IV 65.88\%) and EGD (grade III 6.82\%, IV 60.24\%) test sets.

For the 1p/19q subset, the age of subjects in the internal test set was significantly higher than those in the training set (56 years vs.48 years; $P = 0.001$). However, the WUSM ($P = 0.828$) and EGD ($P = 0.861$) test sets showed no significant difference. In terms of sex, the internal test set had a similar distribution ($P = 0.932$) compared to the training set. However, the WUSM ($P = 0.004$) and EGD ($P = 0.011$) test sets differed significantly due to a higher prevalence of male subjects. In terms of 1p/19q status, the WUSM ($P = 0.994$) and EGD ($P = 0.135$) test sets showed a similar distribution compared to the training set. However, the internal test set consisted of a significantly lower percentage of 1p/19q codeleted cases compared to the training set (13.68\% vs. 34.77\%; $P < 0.001$). In terms of tumor grade, the internal ($P < 0.001$), WUSM ($P < 0.001$), and EGD ($P < 0.001$) test sets differed significantly compared to the training set.

\subsection{Ablation studies for IDH mutation status prediction} \label{supp_results_ablation_IDH}
\subsubsection{Ablation studies with and without prior knowledge features}
We evaluated the performance of the conventional CNN, CNN+age, CNN+loc, and CNN+age+loc models in predicting IDH status and found that the CNN+age model performed significantly better than other models. In comparison, the rest of the models had a performance drop, especially in terms of recall on the WUSM and EGD sets (Supplementary Table~\ref*{supp_table3}). This was more pronounced in the CNN and CNN+loc models ($P < 0.05$ for drop in recall on both WUSM and EGD), demonstrating the importance of age information in the prediction of IDH status. The CNN+age+loc model had a significant drop in precision in the WUSM ($P = 0.006$) and EGD sets ($P < 0.001$). Overall, the CNN+age model performed the best and was used for the subsequent ablation study.

\subsubsection{Ablation studies between 2.5D and planar models}
Within the CNN+age model, the ablation study involving the planar models (Supplementary Table~\ref*{supp_table4}) showed that capturing volumetric spatial information through a multi-view aggregation step is beneficial. Specifically, we found that the 2.5D model significantly outperformed the axial, coronal, and sagittal models in terms of precision on the WUSM and EGD test sets. There was a minor drop in recall compared to the coronal model. This was found to be insignificant for the internal test set, but significant for WUSM (0.162 decrease; $P < 0.001$) and EGD (0.054 decrease; $P = 0.011$) test sets. However, both the axial and sagittal models were significantly outperformed by the 2.5D model in terms of recall on the EGD set with minor improvements on the internal and WUSM sets as well. Due to this overall better performance, the 2.5D CNN+age model was used for the final IDH mutation classification task. 

\subsection{Misclassification analysis for IDH prediction} \label{supp_results_miscl_IDH}
For the WUSM test set, the model misclassified 5.6\% (14 of 249) IDH-wt cases as IDH-mut. This included 3 anaplastic astrocytomas, 1 oligodendroglioma, 1 anaplastic oligodendroglioma, and 1 anaplastic oligoastrocytoma – all with relatively high OS (median OS 49.39 months, range 32.4-82.4 months), low age at diagnosis (median 37.5 years, range 26-51.75 years) and phenotype consistent with prototypical IDH-mut gliomas (Figure~\ref*{fig4}A – case1). We also found a grade-IV glioblastoma located along the dorsal aspect of the midbrain which has radiographic appearance of a low-grade neoplasm with a relatively high OS of 32.4 months (Figure~\ref*{fig4}A – case2). Of the 20.5\% (18 of 88) IDH-mut cases that were misclassified as IDH-wt, we found 3 glioblastoma cases with relatively low OS (median OS 9.17 months, range 4.37-12.7 months), high age at diagnosis (median 74 years, range 72-77 years)  and phenotype consistent with prototypical IDH-wt gliomas, with irregular enhancements (Figure~\ref*{fig4}A – case3). This was also the case for a gliosarcoma (age 74 years), which had a centrally necrotic mass with peripheral enhancement and vaguely defined tumor boundaries and had a low OS of 11.87 months (Figure~\ref*{fig4}A – case4). Prediction of the gliosarcoma as IDH-wt is also consistent with the 2016 WHO classification system~\cite{louis20162016} which includes gliosarcoma under the umbrella of IDH-wt glioblastomas.

For the EGD dataset, of the 11 IDH-mut cases that were misclassified as IDH-wt, three cases comprised scans of coronal acquisition and had very low off-plane resolutions and one case comprised sagittal acquisition with low off-plane resolutions. On a closer inspection, we found that the model made correct predictions for these cases in the plane of acquisition but misclassified them in the other two planes – leading to an overall incorrect prediction. 

\subsection{Ablation studies for 1p/19q codeletion status prediction} \label{supp_results_ablation_1p19q}
\subsubsection{Ablation studies with and without prior knowledge features} 
We evaluated the performance of the conventional CNN, CNN+age, CNN+loc, and CNN+age+loc models and found that the CNN+loc model performed the best. Compared to this, the rest of the schemes performed as follows (Supplementary Table~\ref*{supp_table5}). The conventional CNN exhibited a significant drop in precision on all three test sets ($P < 0.05$) and recall on the WUSM test set ($P = 0.033$). The CNN+age+loc model had a significant drop in precision and recall ($P < 0.05$) on the EGD set. The CNN+age model showed a minor, statistically insignificant increase in precision for all three datasets (internal: 0.112 increase; $P = 0.502$; WUSM: 0.084 increase, $P = 0.161$; EGD: 0.097 increase, $P = 0.121$) test sets. However, it had a significant drop in recall on the WUSM and EGD test sets. Overall, inclusion of only ‘loc’ information led to a performance improvement. Hence, the CNN+loc model was used for the subsequent ablation study.

\subsubsection{Ablation studies between 2.5D and planar models}
Within the CNN+loc model, the ablation study involving the planar models (Supplementary Table~\ref*{supp_table6}) showed that capturing volumetric spatial information is beneficial. Specifically, we found that the 2.5D model yielded significantly better precision than the axial and sagittal models on all three test sets. Compared to the coronal model, the 2.5D model exhibited a minor drop in precision on the WUSM set (0.058 decrease; $P = 0.233$) but a significant improvement in recall on the EGD set. Overall, the 2.5D model performance was better than the planar models and hence, the 2.5D CNN+loc model was used for the final 1p/19q codeletion classification task. 

\subsection{Interpretation of quantitative metrics for 1p/19q status classification} \label{supp_results_metric_details_1p19q}
This disparity in AUROC and AUPRC for 1p/19q classification in the internal test data can be explained by the severe class-imbalance. The over-represented negative class (i.e., 1p/19q non-codeleted) in this dataset results in a high number of true negative (TN) predictions and in turn, a low false positive rate (FP/FP+TN). This leads to an overly optimistic performance in terms of AUROC. For a similar reason, the accuracy values are also affected and need to be interpreted with caution. On the other hand, the precision-recall curve only addresses the performance on the positive samples (i.e., 1p/19q codeleted) of the dataset and is not affected by the high number of negative samples. This also makes AUPRC a more reliable metric than AUROC under class-imbalance~\cite{saito2015precision}.

\newpage

\begin{table}[htbp]
\centering

\caption{Statistical comparison of performance between proposed model and Voort-CNN and Choi-CNN for prediction of IDH mutation status.}
\label{supp_table1}
\adjustbox{max width = \textwidth}{
\begin{tabular}{l|c|c|c|c|c|c|c|c|c}
\hline
          & Precision & Difference & P-value          & Recall & Difference & P-value & AUROC & Difference & P-value          \\ \hline
Voort-CNN & 0.453     & -0.391     & \textless{}0.001 & 0.574  & -0.22      & 0.004   & 0.755 & -0.113     & 0.002            \\
Choi-CNN  & 0.438     & -0.393     & \textless{}0.001 & 0.644  & -0.149     & 0.028   & 0.704 & -0.17      & \textless{}0.001 \\ \hline
\end{tabular}}

\vspace{0.5cm} 

\caption{Statistical comparison of performance between proposed model and Voort-CNN for prediction of 1p/19q codeletion status.}
\label{supp_table2}
\adjustbox{max width = \textwidth}{
\begin{tabular}{l|c|c|c|c|c|c|c|c|c}
\hline
          & Precision & Difference & P-value          & Recall & Difference & P-value & AUROC & Difference & P-value          \\ \hline
Voort-CNN & 0.667                          & -0.06                           & 0.827                        & 0.045                       & -0.682                          & \textless{}0.001             & 0.673                      & -0.06                           & 0.371                        \\ \hline
\end{tabular}}

\vspace{0.5cm} 

\caption{Statistical comparison of performance between CNN, CNN+loc, CNN+age+loc models with proposed model (CNN+age) for prediction of IDH mutation status.}
\label{supp_table3}
\centering
\adjustbox{max width = \textwidth}{
\begin{threeparttable}
\begin{tabular}{l|c|c|c|c|c|c|c|c|c}
\hline
            & Precision & Difference & P-value          & Recall & Difference & P-value          & AUROC & Difference & P-value          \\ \hline
Test        &           &            &                  &        &            &                  &       &            &                  \\ 
~~~~~CNN         & 1         & 0.056      & 0.303            & 0.75   & -0.1       & 0.157            & 0.926 & 0.001      & 0.978            \\ 
~~~~~CNN+loc     & 0.579     & -0.365     & \textless{}0.001 & 0.55   & -0.3       & 0.014            & 0.738 & -0.187     & 0.004            \\ 
~~~~~CNN+age+loc & 0.889     & -0.055     & 0.530            & 0.8    & -0.05      & 0.564            & 0.912 & -0.013     & 0.657            \\ \hline
WUSM        &           &            &                  &        &            &                  &       &            &                  \\ 
~~~~~CNN         & 0.873     & 0.046      & 0.254            & 0.632  & -0.156     & \textless{}0.001 & 0.854 & -0.019     & 0.280            \\ 
~~~~~CNN+loc     & 0.456     & -0.371     & \textless{}0.001 & 0.612  & -0.176     & \textless{}0.001 & 0.73  & -0.143     & \textless{}0.001 \\ 
~~~~~CNN+age+loc & 0.733     & -0.094     & 0.006            & 0.741  & -0.047     & 0.248            & 0.849 & -0.024     & 0.187            \\ \hline
EGD         &           &            &                  &        &            &                  &       &            &                  \\ 
~~~~~CNN         & 0.914     & 0.007      & 0.770            & 0.858  & -0.068     & 0.025            & 0.939 & 0.006      & 0.661            \\ 
~~~~~CNN+loc     & 0.623     & -0.284     & \textless{}0.001 & 0.736  & -0.19      & \textless{}0.001 & 0.824 & -0.109     & \textless{}0.001 \\ 
~~~~~CNN+age+loc & 0.79      & -0.117     & \textless{}0.001 & 0.865  & -0.061     & 0.039            & 0.909 & -0.024     & 0.049            \\ \hline
\end{tabular}
\begin{tablenotes}
      \small
      \item Abbreviations: WUSM, Washington University School of Medicine; EGD, Erasmus Glioma Database; AUROC, area under the receiver operating characteristic curve; CNN, conventional CNN; CNN+loc, CNN with tumor location information; CNN+age CNN with patient age information; CNN+age+loc, CNN with patient age and tumor location information.
    \end{tablenotes}
  \end{threeparttable}
}

\vspace{0.5cm} 

\caption{Statistical comparison of performance between axial, coronal, sagittal models with proposed model for prediction of IDH mutation status.}
\label{supp_table4}
\adjustbox{max width = \textwidth}{
\begin{threeparttable}
\begin{tabular}{l|c|c|c|c|c|c|c|c|c}
\hline
         & Precision & Difference & P-value          & Recall & Difference & P-value          & AUROC & Difference & P-value          \\ \hline
Test     &           &            &                  &        &            &                  &       &            &                  \\ 
~~~~~axial    & 1         & 0.056      & 0.304            & 0.5    & -0.35      & 0.008            & 0.896 & -0.029     & 0.419            \\ 
~~~~~coronal  & 0.514     & -0.43      & \textless{}0.001 & 0.95   & 0.1        & 0.157            & 0.799 & -0.126     & 0.007            \\ 
~~~~~sagittal & 0.762     & -0.182     & 0.074            & 0.8    & -0.05      & 0.317            & 0.88  & -0.045     & 0.276            \\ \hline
WUSM     &           &            &                  &        &            &                  &       &            &                  \\ 
~~~~~axial    & 0.732     & -0.097     & 0.023            & 0.605  & -0.186     & \textless{}0.001 & 0.841 & -0.033     & 0.119            \\ 
~~~~~coronal  & 0.347     & -0.482     & \textless{}0.001 & 0.953  & 0.162      & \textless{}0.001 & 0.808 & -0.066     & 0.004            \\ 
~~~~~sagittal & 0.508     & -0.321     & \textless{}0.001 & 0.756  & -0.035     & 0.083            & 0.787 & -0.087     & \textless{}0.001 \\ \hline
EGD      &           &            &                  &        &            &                  &       &            &                  \\ 
~~~~~axial    & 0.815     & -0.092     & \textless{}0.001 & 0.682  & -0.244     & \textless{}0.001 & 0.888 & -0.045     & \textless{}0.001 \\ 
~~~~~coronal  & 0.505     & -0.402     & \textless{}0.001 & 0.98   & 0.054      & 0.011            & 0.849 & -0.084     & \textless{}0.001 \\ 
~~~~~sagittal & 0.743     & -0.164     & \textless{}0.001 & 0.899  & -0.027     & 0.046            & 0.89  & -0.043     & 0.003            \\ \hline
\end{tabular}
\begin{tablenotes}
      \small
      \item Abbreviations: WUSM, Washington University School of Medicine; EGD, Erasmus Glioma Database; AUROC, area under the receiver operating characteristic curve.
    \end{tablenotes}
  \end{threeparttable}
}
\end{table}

\begin{table}[!t]
\centering
\caption{Statistical comparison of performance between CNN, CNN+age, CNN+age+loc models with proposed model (CNN+loc) for prediction of 1p/19q codeletion status.}
\label{supp_table5}
\adjustbox{max width = \textwidth}{
\begin{threeparttable}
\begin{tabular}{l|c|c|c|c|c|c|c|c|c}
\hline
            & Precision & Difference & P-value          & Recall & Difference & P-value & AUROC & Difference & P-value          \\ \hline
Test        &           &            &                  &        &            &         &       &            &                  \\ 
~~~~~CNN         & 0.333     & -0.255     & 0.050            & 0.5    & -0.125     & 0.317   & 0.776 & 0.046      & 0.450            \\ 
~~~~~CNN+age     & 0.7       & 0.112      & 0.502            & 0.438  & -0.187     & 0.257   & 0.451 & -0.279     & 0.064            \\ 
~~~~~CNN+age+loc & 0.3       & -0.288     & 0.025            & 0.375  & -0.25      & 0.102   & 0.55  & -0.18      & 0.100            \\ \hline
WUSM        &           &            &                  &        &            &         &       &            &                  \\ 
~~~~~CNN         & 0.455     & -0.283     & \textless{}0.001 & 0.615  & -0.123     & 0.033   & 0.674 & -0.079     & 0.060            \\ 
~~~~~CNN+age     & 0.822     & 0.084      & 0.161            & 0.569  & -0.169     & 0.022   & 0.654 & -0.099     & 0.057            \\ 
~~~~~CNN+age+loc & 0.667     & -0.071     & 0.178            & 0.677  & -0.061     & 0.206   & 0.681 & -0.072     & 0.075            \\ \hline
EGD         &           &            &                  &        &            &         &       &            &                  \\ 
~~~~~CNN         & 0.481     & -0.255     & \textless{}0.001 & 0.694  & -0.063     & 0.225   & 0.795 & -0.046     & 0.152            \\ 
~~~~~CNN+age     & 0.833     & 0.097      & 0.121            & 0.571  & -0.186     & 0.005   & 0.637 & -0.204     & \textless{}0.001 \\ 
~~~~~CNN+age+loc & 0.581     & -0.155     & 0.010            & 0.614  & -0.143     & 0.018   & 0.648 & -0.193     & \textless{}0.001 \\ \hline
\end{tabular}
\begin{tablenotes}
      \small
      \item Abbreviations: WUSM, Washington University School of Medicine; EGD, Erasmus Glioma Database; AUROC, area under the receiver operating characteristic curve; CNN, conventional CNN; CNN+loc, CNN with tumor location information; CNN+age CNN with patient age information; CNN+age+loc, CNN with patient age and tumor location information.
    \end{tablenotes}
  \end{threeparttable}
}

\vspace{0.5cm} 

\caption{Statistical comparison of performance between axial, coronal, sagittal models with proposed model for prediction of 1p/19q codeletion status.}
\label{supp_table6}
\adjustbox{max width = \textwidth}{
\begin{threeparttable}
\begin{tabular}{l|c|c|c|c|c|c|c|c|c}
\hline
         & Precision & Difference & P-value          & Recall & Difference & P-value & AUROC & Difference & P-value \\ \hline
Test     &           &            &                  &        &            &         &       &            &         \\ 
~~~~~axial    & 0.333     & -0.223     & 0.016            & 0.562  & -0.063     & 0.564   & 0.602 & -0.12      & 0.131   \\ 
~~~~~coronal  & 0.444     & -0.112     & 0.285            & 0.5    & -0.125     & 0.317   & 0.67  & -0.052     & 0.500   \\ 
~~~~~sagittal & 0.364     & -0.192     & 0.021            & 0.75   & 0.125      & 0.317   & 0.803 & 0.081      & 0.438   \\ \hline
WUSM     &           &            &                  &        &            &         &       &            &         \\ 
~~~~~axial    & 0.616     & -0.122     & 0.006            & 0.692  & -0.046     & 0.366   & 0.71  & -0.044     & 0.308   \\ 
~~~~~coronal  & 0.796     & 0.058      & 0.233            & 0.662  & -0.076     & 0.059   & 0.73  & -0.024     & 0.494   \\ 
~~~~~sagittal & 0.587     & -0.151     & \textless{}0.001 & 0.831  & 0.093      & 0.058   & 0.755 & 0.001      & 0.984   \\ \hline
EGD      &           &            &                  &        &            &         &       &            &         \\ 
~~~~~axial    & 0.56      & -0.164     & \textless{}0.001 & 0.708  & -0.056     & 0.206   & 0.75  & -0.092     & 0.017   \\ 
~~~~~coronal  & 0.638     & -0.086     & 0.072            & 0.611  & -0.153     & 0.005   & 0.79  & -0.052     & 0.126   \\ 
~~~~~sagittal & 0.536     & -0.188     & \textless{}0.001 & 0.819  & 0.055      & 0.248   & 0.831 & -0.011     & 0.692   \\ \hline
\end{tabular}
\begin{tablenotes}
      \small
      \item Abbreviations: WUSM, Washington University School of Medicine; EGD, Erasmus Glioma Database; AUROC, area under the receiver operating characteristic curve.
    \end{tablenotes}
  \end{threeparttable}
}
\end{table}


\end{document}